\documentclass{aa}  
\usepackage[switch]{lineno}

\usepackage{multirow}
\usepackage{natbib}
\usepackage[english]{babel}
\usepackage{graphicx}
\usepackage{txfonts}
\usepackage{booktabs}
\usepackage{multicol}
\usepackage{makecell}
\usepackage{wasysym}
\usepackage{amsmath}
\usepackage{booktabs}
\usepackage{threeparttable} 
\usepackage{hyperref}
\hypersetup{
  colorlinks   = true,    
  urlcolor     = blue,    
  linkcolor    = blue,    
  citecolor    = blue     
}

\newcommand{\msun}{M$_\odot$}

\defcitealias{paperI}{paper\ I}

\begin{document} 

   \title{Tracing the evolution of brightest galaxies and diffuse light in galaxy groups}


   \author{B.\ Bilata-Woldeyes
          \inst{1}\fnmsep\thanks{Email: betelehem@iaa.csic.es}, 
          J.\ D.\ Perea\inst{1}
          \and
          J.\ M.\ Solanes\inst{2}
          }

   \institute{Instituto de Astrof\'\i sica de Andaluc\'\i a (IAA–CSIC), Glorieta de la Astronom\'\i a, s/n, E-18008 Granada, Spain
         \and
             Departament de F\'\i sica Qu\`antica i Astrof\'\i sica and Institut de Ci\`encies del Cosmos (ICCUB), Universitat de Barcelona. C.\ Mart\'{\i}  i Franqu\`es, 1, E-08028 Barcelona, Spain
             }

   \date{v1.21 last revised Oct. 21, 2025); first submitted in July 31, 2025; accepted October 27, 2025}

  \abstract
   {We present the second study based on a suite of 100 cosmologically motivated, controlled $N$-body simulations designed to advance the understanding of the role of purely gravitational dynamics in the early formation of low-mass galaxy groups ($\sim\!1$--$5 \times 10^{13}\,$\msun). In this work, we investigate the temporal evolution of key indicators of dynamical relaxation, with particular emphasis on the secular growth of the diffuse intragroup light (IGL), the four major group galaxies, and the mass distributions of their progenitors. We also assess the diagnostic power of several magnitude gaps between top-ranked galaxies as proxies for dynamical age. As in our previous study, we compare outcomes from three group classes defined by the number of brightest group galaxies (BGGs) present at the end of the simulations: single-BGG, double-BGG, and non-BGG systems.\\\noindent
   We find that the early assembly of galaxy groups is consistent with a stochastic Poisson process at an approximately constant merger rate. Various dynamical diagnostics -- including galaxy pairwise separations, velocity dispersions, and the offset of the first-ranked galaxy from the group barycentre -- indicate that single-BGG groups evolve more rapidly towards virialisation than double- and especially non-BGG systems. We further find that first-ranked group members and the IGL, despite their intertwined origins, follow distinct growth histories, with the IGL assembled from a more numerous and systematically lower-mass population than the central object. This distinction is particularly pronounced in non-BGG systems, where about one third of the IGL originates from small galaxies, each contributing less than $5\%$ to this component. Among the tested magnitude gaps, the difference between the first- and fourth-ranked galaxies, $\Delta{\cal{M}}_{\rm{4-1}}$, proves a more robust indicator of dynamical age for low-mass groups than the conventional $\Delta{\cal{M}}_{\rm{2-1}}$ gap. The $\Delta{\cal{M}}_{\rm{5-1}}$ and $\Delta{\cal{M}}_{\rm{6-1}}$ gaps also perform well and may be preferable in certain contexts.}


\titlerunning{Evolution of brightest group galaxies and diffuse light}
\authorrunning{B.\ Bilata-Woldeyes et al.}
\maketitle

\section{Introduction}\label{Intro}

Galaxy groups are the most common galaxy associations in the universe and the building blocks of larger structures such as clusters and superclusters of galaxies \citep[e.g.][]{2000ASPC..209..434D, Paul+2017}. Despite their modest masses, typically ranging from $10^{13}$ to $10^{14}\,$\msun, groups play a crucial role in galaxy evolution by fostering interactions and mergers between galaxies that shape their morphology and star formation histories \citep[e.g.][]{2001ASPC..240..547Z, McGee+2009}. A particularly intriguing subset of these systems is low-mass groups at early stages of their dynamical evolution, which offer valuable insights into the processes governing hierarchical structure formation.

A defining feature in a good number of these systems is the presence of a dominant brightest group galaxy (BGG\footnote{\label{fn:clusters}For galaxy clusters, the equivalent terms are BCG for the brightest cluster galaxy and ICL for intracluster light.}) typically identified in the optical or near-infrared bands, where luminosity correlates well with stellar mass \citep{2007ApJ...671..153Y, 2014A&A...566A.140G}. BGGs are often located near the system's barycentre and/or at the peak of the X-ray emission, underscoring their dynamical centrality within the group \citep{2004ApJ...617..879L, 2011ApJ...742..125G, 2019MNRAS.483.3545G}. However, this is not always the case, particularly in dynamically young systems, where the BGG may be offset from the group's potential minimum \citep{2011MNRAS.410..417S, 2024A&A...681A..91E}. 

Galaxy groups also exhibit varying amounts of intragroup light (IGL\textsuperscript{\ref{fn:clusters}}), an extended, low-surface-brightness stellar component that resides between galaxies and that is not gravitationally bound to any of them. This diffuse light originates primarily from stars stripped from group members through violent disruptive gravitational interactions \citep{2005MNRAS.358..949Z, 2006ApJ...648..936R, 2009ApJ...699.1518R, 2013ApJ...778...14G, 2018MNRAS.479..932C, 2019A&A...622A.183J}, although a portion may also form in situ through the compression of the intragroup gas \citep{2022ApJ...930...25B, 2023MNRAS.521..800M, 2024OJAp....7E.111A}, particularly during the early assembly phase of the groups \citep{2003ApJ...589..752G, 2010MNRAS.406..936P, 2016MNRAS.461..321S}. Most of the IGL is usually found concentrated around the BGG \citep{2015IAUGA..2247903M, 2021Galax...9...60C, 2022NatAs...6..308M}, suggesting that both components share a common origin driven by the evolutionary history of the host group. Studying the co-formation and co-evolution of the BGG and IGL thus provides valuable constraints on the assembly of galaxy groups, the impact of environment on galaxy evolution, and the distribution of cold dark matter (CDM) on supergalactic scales \citep{2019MNRAS.482.2838M, 2020ApJ...901..128C, 2022ApJS..261...28Y}. 

However, using the observational properties of these two components to infer group dynamical age remains challenging due to their proximity in projected phase space, which makes it difficult to define clear boundaries between them \citep{2011ApJ...732...48R, 2021ApJ...910...45M, 2022MNRAS.514.3082M}, incomplete membership information, and the fact that they evolve at different rates and timescales \citep{2017ApJ...846..139M, 2018MNRAS.479..932C}.

A promising avenue to estimate the degree of dynamical relaxation of galaxy groups involves the use of magnitude gaps, defined as the difference in magnitude in a given photometric band sensitive to the old stellar population, between the most luminous galaxy and one of its next-ranked companions within a fixed system's radius. The appeal of such metrics as tracers of the evolutionary history of galaxy overdensities lies in the close link between the stellar mass growth of the central galaxy -- driven primarily by mergers within transient substructures during the group’s pre-virialisation phase -- and the assembly of the host halo \citep[e.g.][]{2016MNRAS.461..321S}. In contrast, satellite galaxies are typically thought to either eventually merge with the central object or remain largely unaffected by this hierarchical growth process \citep{2009MNRAS.395.1376W, 2019MNRAS.485.2287B}, although, as this work will show, this assumption may not always hold. Coupled with their ease of measurement, these properties make magnitude gaps one of the most widely employed diagnostics of the dynamical state of galaxy systems, especially in large datasets, without the need for detailed dynamical modelling.

In the context of galaxy groups, the first application of a magnitude gap was introduced by \citet{Ponman+1994}, who proposed the existence of fossil groups --systems dominated by a single, bright elliptical galaxy -- based on an extreme gap of $\Delta{\cal{M}}_{2-1}=2$ mag in the Cousins $R$-band, measured between the BGG and the largest satellite, within a projected radius of $0.5\,R_{\rm{vir}}$\footnote{Some subsequent studies adopted a fixed physical radius of 500 kpc for identifying fossil candidates \citep[e.g.][]{Santos+2007}.} \citetext{see also \citealt{Jones+2003} and \citealt{D'Onghia+2005}}. Later, \citet{Milosavljevic+2006} generalized the application of this criterion by using both $\Delta{\cal{M}}_{2-1}$ and $\Delta{\cal{M}}_{3-1}$ to estimate the dynamical age of galaxy clusters identified in the Sloan Digital Sky Survey. 

More recently, \citet{Dariush+2010}, analysing groups and clusters of galaxies from the Millennium Simulation, proposed a refined photometric criterion based on a gap of $\Delta{\cal{M}}_{4-1}> 2.5$ mag in the Sloan $r$-band measured between the BGG and the fourth-ranked galaxy. They showed that this alternative definition more reliably identifies early-formed systems compared to the traditional $\Delta{\cal{M}}_{2-1}$ gap, particularly in those that may have experienced recent infall of bright satellites. Since then, $\Delta{\cal{M}}_{4-1}$ has gained widespread adoption not only as a fossil group identifier, but also as a practical, accessible estimator for dynamical maturity across all kind of galaxy systems in both observational and simulation-based studies \citetext{e.g.\ \citealt{Harrison+2012, Zarattini+2014, Kanagusuku+2016, Golden-Marx+2018, Farahi+2020, Golden-Marx+2022},~\citeyear{Golden-Marx+2025}}.

Despite the popularity of these simple, measurable statistics as proxies for dynamical state, particularly in fossil groups, their broader use in systems with more intricate merger histories has received surprisingly little critical scrutiny by the astronomical community. To our knowledge, there has been no comprehensive assessment of whether these metrics retain diagnostic validity outside the fossil group regime.

To address this issue in a controlled theoretical setting, as well as to further our understanding of the role played by purely gravitational dynamics in the formation and evolution of low-mass galaxy groups, we present this second study based on a suite of 100 cosmologically motivated, dissipationless $N$-body simulations that trace the early gravitational collapse of galaxy groups with total masses in the range $\sim 1$--$5$ $\times\,10^{13}\,$\msun. In our first paper, \citeauthor{paperI} (\citeyear{paperI}; hereafter \citetalias{paperI}), we introduced the simulation setup and examined how various physical parameters of galaxy groups influence the formation of ex situ IGL during the initial stages of their assembly. In this second instalment, we extend that analysis by tracking the evolution of several indicators of dynamical relaxation, with a particular focus on assessing the diagnostic power of different magnitude gaps involving the brightest group galaxies.

This paper is organized as follows. In Sect.~\ref{sec:group_simulations}, we review the numerical framework used to model pre-virialised galaxy groups and their constituent galaxies, along with the procedure employed to separate galaxies from the IGL component that forms during the simulations. We also explain how we divide the group sample into three outcome classes, single-BGG, double-BGG, and non-BGG systems, to determine how the dynamical markers whose evolutionary histories are examined in Sect.~\ref{sec:non-photometric_properties} evolve under different assembly scenarios. Sect.~\ref{sec:photometric_properties} presents a detailed analysis of the IGL formation history across group classes, highlighting the factors and progenitor galaxies driving its growth, and examines a range of IGL-based indicators of group dynamics. A similar group-by-group approach is applied in Sect.~\ref{sec:top-ranked-galaxies}, where we trace the evolution of the four top-ranked group galaxies and also identify their main progenitors. Finally, in Sect.~\ref{sec:evaluating_mgap} we use correlations between various magnitude gaps defined from the most luminous group galaxies and the stellar mass fractions locked in both the first-ranked galaxy and the IGL, as well as their dependence on the assembly histories of some of the most massive satellites, to identify the gaps that better perform as age indicators in dynamically complex systems like our forming groups. The main findings of this work are summarised in Sect.~\ref{sec:conclusions}.

For this study, we adopt the same flat $\Lambda$CDM cosmology used in \citet{2016MNRAS.461..321S}, which is consistent with the Wilkinson Microwave Anisotropy Probe 5-yr results \citep{2009ApJS..180..330K} that show a matter density parameter $\Omega_{\mathrm{m,0}} = 1 - \Omega_{\Lambda,0}= 0.26$ and a Hubble-Lema\^\i tre constant of $H_0 = 100\,h\;\mbox{km}\,\mbox{s}^{-1}\,\mbox{Mpc}^{-1}$, with $h=0.72$.

\section{Galaxy group simulations} \label{sec:group_simulations}
\subsection{Group and galaxy modelling} \label{ssec:models}

Our simulations of the hierarchical growth of galaxy groups during the epoch leading up to their first gravitational collapse are based on a computationally efficient $N$-body framework designed to model pre-virialised galaxy aggregations. This model, developed by some of us, has previously been used to delve into the optical properties of present-day galaxy groups and their brightest members \citep{2016MNRAS.461..321S}, as well as to explore the impact of dissipationless hierarchical merging on the mass Fundamental Plane defined by massive early-type galaxies \citep{2016MNRAS.461..344P}. In this approach, galaxy groups are initially represented by uniform, isolated spherical top-hat density perturbations that expand with the Hubble flow, subsequently reach turnaround, and ultimately undergo full non-linear collapse. Neglecting pressure gradients, the evolution of each overdensity follows the dynamics of a closed Friedmann universe, starting from a radius
\begin{equation}\label{overdensity}
R_{\mathrm{grp}}(t_{\rm ini})=\left[\frac{3 M_{\mathrm{grp}}}{4\pi\rho_{\rm crit}(t_{\rm ini})\Omega_{\mathrm m}(t_{\rm ini})(1+\delta_{\rm ini})}\right]^{1/3}\;,
\end{equation}
where the critical density, $\rho_{\rm crit}(t_{\rm ini})$, and the mass density parameter, $\Omega_{\rm m}(t_{\rm ini})$, refer to the unperturbed adopted background cosmology at cosmic time $t_{\rm ini}$, corresponding to an initial redshift $z_{\rm ini}=3$. The value of the initial perturbation amplitude, $\delta_{\rm ini}>0$, is calibrated such that a perfectly homogeneous spherical overdensity with the same total mass as the group, $M_{\rm{grp}}$, would collapse to a point by the final redshift $z_{\rm fin} = 0$ of the runs. This setup ensures that all the simulated groups evolve on cosmologically consistent timescales. 

For their part, the member galaxies are modelled as rotating extended CDM spheres with Navarro-Frenk-White density profiles \citep{1997ApJ...490..493N}, whose total (virial) masses are randomly drawn from a Schechter-like mass function (MF) of asymptotic slope $\alpha=-1.0$ and characteristic mass $M^*=10^{12}\,h^{-1}\,$\msun\ -- corresponding to the halo mass of a Milky Way-type galaxy \citep[e.g.][]{2010MNRAS.406..896B, Watkins+2019, Deason+2020, Shen+2022} --, truncated below $0.05\,M^*$. The total galaxy mass sets the scale for other global properties, including the CDM halo spin and concentration, as well as the mass of the baryonic core, which is fixed at $5\%$ of the total mass \citep[e.g.][]{Governato+2007, Sales+2009, Dai+2010, 2010A&A...516A...7D}. Throughout the manuscript, we refer to this component as the luminous or stellar component, although it formally also includes the less important contributions from the gas and dust in the interstellar medium. This means that in our collisionless simulations, the mass of this component is not affected by the star formation histories of galaxies. For galaxies with masses above $0.1\,M^*$ the luminous component is configured either as an exponential disc plus a \citet{1990ApJ...356..359H} spheroidal bulge or as a single spheroid, with structural and dynamical properties that align with the main scaling laws defined by real objects of the same morphology in the local universe. Morphologies are assigned via a Monte Carlo method that assumes an initial late-type fraction of $0.80$ on average, consistent with the morphological distribution of the JWST galaxies \citep{Lee+2024}. Galaxies below $0.1\,M^*$ are modelled, for simplicity, as purely spheroidal systems. Each galaxy model, regardless of its total mass, is constructed with an equal number of luminous and dark bodies, with individual masses of $\sim 2.3\times 10^5\,$\msun\ and $\sim 6.6\times 10^6\,$\msun, respectively. The Plummer-equivalent softening length is fixed at 30 pc for luminous (baryonic) particles, while for CDM particles it scales with the square root of their mass, ensuring uniform maximum inter-particle gravitational forces across all simulations. After establishing the central distributions of the progenitor galaxies, the total masses and taper radius of their CDM haloes are scaled down to values consistent with the initial redshift of the simulations, while the stellar cores are left unchanged.

The initial number of galaxies per group, $N_{\rm{gal}}(z_{\rm ini})$, is determined to match the subhalo abundance statistics of typical group-sized $\Lambda$CDM haloes \citep[e.g.][]{2011MNRAS.410.2309G}. We treat $N_{\rm{gal}}(z_{\rm ini})$ as a Poisson random variable with mean and variance $\lambda$, given by
\begin{equation}\label{N_memb}
\lambda\equiv N_{\rm bright}(z_{\rm ini})\frac{M_{\rm grp}}{10M^{\rm su}}\frac{\Phi(\alpha,M_{\rm min})}{\Phi(\alpha,0.5M^*)}\;,
\end{equation}
where $M_{\rm min}=0.05M^*$ is the low-mass cutoff adopted for the Schechter galaxy mass function, $N_{\rm{bright}}(z_{\rm ini})=5$ is the initial number of massive (luminous) galaxies (i.e.\ with total mass $\geq 0.5M^*$) -- which given the typical group dimensions implies an average density of five large galaxies per Mpc$^3$ --, $\Phi(\alpha,x)$ is the cumulative Schechter MF, and $M^{\rm su}=M^*$ is the simulation mass unit. This setup produces galaxy counts that scale with group mass, yielding values of $N_{\rm{gal}}(z_{\rm ini})$ that for this specific suite of 100 group simulations range between 20 and 69, with a median of 31. The quartiles of the distribution of total group masses are $(Q_1,Q_2,Q_3)=(1.63,2.13,2.76)\,\times\,10^{13}\,$\msun, while those of the stellar masses, $M_\star$, are $(6.87, 8.43, 11.2)\,\times\,10^{11}\,$\msun. The total-mass-to-stellar-mass-ratios of the groups span from 15 to 43, with a median of 24.

At $z_{\rm ini}$, galaxy centres of mass are randomly distributed within each group volume, ensuring that their CDM haloes do not overlap, and their initial velocities set by the local Hubble flow. The intragroup background is uniformly populated with CDM particles -- with masses identical to those used in the galaxy models -- until the total target mass of the group is reached. During the early phase of the simulations -- approximately the first third of the total runtime of $\sim 11.5$ Gyr, encompassing the expansion and turnaround stages --, group members gradually decouple from the Hubble flow as self-gravity slows and eventually reverses the system's expansion. This phase allows galaxies to reach dynamical equilibrium within the gravitational potential of their host group and to develop realistic peculiar velocities and higher-order spatial correlations before the onset of the non-linear, merger-dominated collapse phase. To track the full evolution of each system, a total of $120$ snapshots were generated, uniformly spaced in cosmic time from $z_{\rm ini} = 3$ to $z_{\rm fin} = 0$.

\subsection{Galaxy and IGL identification across snapshots} \label{ssec:identification}

To isolate the diffuse IGL from galaxy-bound luminous particles in each simulation snapshot, we applied a surface brightness (SB) threshold to the synthetic images, thereby separating low-density regions from the denser galactic cores. Following the approach of \citet{2005ApJ...631L..41M}, \citet{2011ApJ...732...48R}, and \citet{2014MNRAS.437..816C}, among others, we adopted a SB cutoff of $26.5\,V$-mag arcsec$^{-2}$ to identify regions associated with the IGL. This observational cutoff was first converted into physical units of solar luminosity per square parsec and subsequently translated into stellar mass using an assumed mass-to-light ratio of $5\,\Upsilon_{V,\odot}$ \citep[e.g.][]{2011ApJ...732...48R}, yielding a projected stellar mass density threshold of $4.5\,$\msun$\,\mbox{pc}^{-2}$. This criterion was applied to all stellar particles in three orthogonal (independent) projections for each simulation snapshot. Both 2D and 3D stellar densities were estimated using a KD-tree algorithm implemented in Fortran by Li Dong\footnote{\url{https://github.com/dongli/fortran-kdtree}}, with local densities computed based on the distance to the $50$-th nearest neighbour.

Once IGL particles were identified and removed from the simulated group images, the remaining stellar particles were used to reconstruct the galaxy population. This was accomplished with the aid of the centroid-based Mean-Shift unsupervised clustering algorithm \citep[see e.g.][]{Carreira2015}, which determines the 3D positions of galaxy centroids without requiring a priori knowledge of their number. The algorithm works by iteratively shifting each particle toward the local maximum of a density function, defined within a kernel window of fixed bandwidth, effectively tracing the particle to the nearest density peak. This process continues until convergence is reached. A subsequent post-processing step filters out nearby duplicate peaks, yielding a clean set of galaxy centroids. The number of galaxies in a given snapshot was thus determined by the final count of unique density maxima identified through this procedure. To ensure robust identification, both the kernel bandwidth and centroid estimates were computed using only the luminous particles in the densest regions. This strategy reduces ambiguities in galaxy identification caused by the overlap of the extended outer stellar envelopes, which can become particularly problematic in snapshots dominated by advanced mergers. Tests exploring a range of local density thresholds during the first half of our simulation timeline revealed that the galaxy-finding algorithm performs best when restricted to luminous particles with 3D local densities above the 65th percentile of their ranked density distribution within each snapshot. This adaptive threshold effectively isolates the densest stellar regions, corresponding to galactic cores, and provides a robust basis for centroid estimation. We further verified that, under this configuration, the galaxy positions and counts identified by the Mean-Shift clustering algorithm closely match those obtained with the widely used source detection tool \textsc{SExtractor} \citep{1996A&AS..117..393B}, when applied to any of the three orthogonal projections of the simulations snapshots.

\begin{figure}
\centering
   \includegraphics[width=9cm, height=6.5cm]{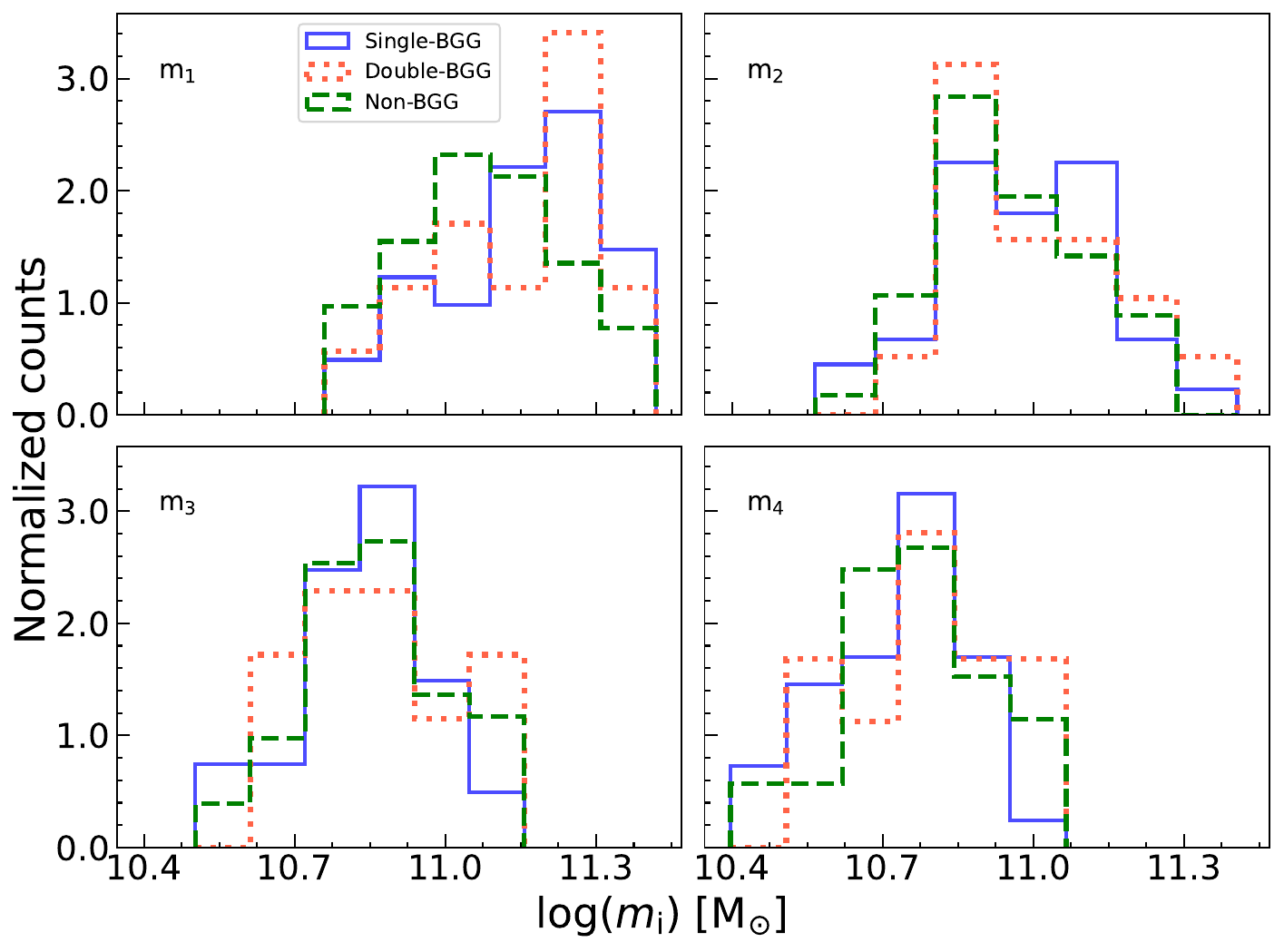}
    \caption{Luminous mass distribution of the four most massive galaxies at the start of the simulations, categorized by group class. Group classes are coded as follows: single-BGG systems (blue solid line), double-BGG systems (orange dotted line), and non-BGG systems (green dashed line).}
    \label{fig:init_mgal_dist}
\end{figure}

\begin{figure}
\centering
 \includegraphics[width=9cm, height=6.88cm]{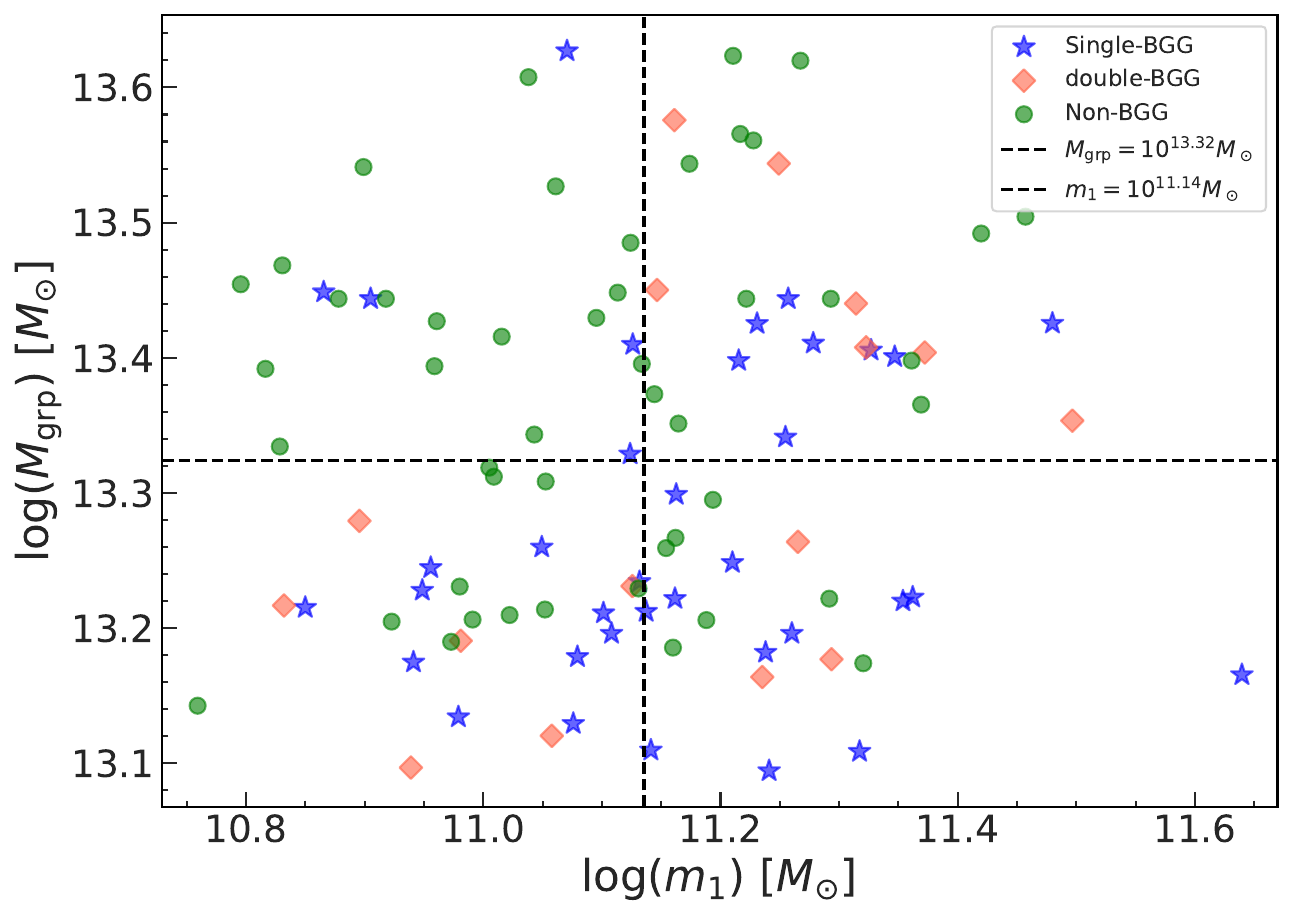}
    \caption{Total group mass, $M_{\rm grp}$, as a function of the mass of the first-ranked galaxy, $m_1$, at $z_{\rm ini}=3$. The horizontal dashed line marks the median total mass of all groups, while the vertical dashed line indicates the median mass of the first-ranked galaxies. The absence of any obvious clustering or segregation in the data within the quadrants delineated by these dividing lines suggests no significant bias in the initial conditions. Group classes are coded as follows: single-BGG systems (blue stars), double-BGG systems (orange diamonds), and non-BGG systems (green circles).} 
    \label{fig:m1_mtot}
\end{figure}

After identifying the centroids of all galaxies in the group images, we assigned the luminous particles not classified as IGL to individual galaxies to determine their total stellar masses. This was accomplished through a recursive procedure based on an ad-hoc measure of the linking strength, $f_i$, between each particle and every galaxy centroid, which we defined through the relation
\begin{equation}\label{force}
f_i\equiv\frac{m_i}{d_i^4}\;,     
\end{equation}
where $m_i$ denotes the stellar mass of galaxy $i$ and $d_i$ is the 3D distance between the particle and the corresponding galaxy centroid. In the initial step, $m_i$ was approximated using the luminous mass contained within the galaxy’s densest core region -- specifically, the top $35\%$ of luminous particles with the highest local densities, consistent with the method used for centroid identification. Next, the remaining $65\%$ of non-IGL particles with the lowest local densities were assigned to the galaxy whose centroid yielded the maximum value of $f_i$, thereby completing the first mass allocation. Following this, all linking strengths were recalculated using updated values of $m_i$, now defined as the total stellar mass of galaxy $i$ obtained by summing the masses of all non-IGL particles currently assigned to it. A  new reassignment of the low-density non-IGL particles was then carried out based on the revised linking strengths. This last part of the process was iterated -- recomputing linking strengths and reassigning particles -- until convergence was reached when no further changes occurred in the particle-to-galaxy assignments. Finally, galaxies were ranked in descending order according to their final stellar masses. 

The inconsistency inherent in using a 2D surface density threshold to identify the IGL alongside a 3D linking-length-based method to estimate galaxy stellar masses has a negligible impact on our results. In each run, the typical discrepancy between the total mass of stellar particles and the sum of the luminous mass assigned to galaxies and the IGL remains around $1\%$, with the maximum deviation never exceeding $3.5\%$. Further details on the modelling of groups and their member galaxies, as well as on the methodology used to separate the luminous components, can be found in \citetalias{paperI}, as well as in \citet{2016MNRAS.461..321S} and \citet{2010A&A...516A...7D}.

\subsection{Group classification} \label{ssec:group_classes}

As demonstrated in \citetalias{paperI}, by the end of the simulations, our groups naturally segregate into three distinct categories defined by the prominence of the first-ranked galaxy\footnote{In observational terms, a first-ranked galaxy ($m_1$) is the brightest member of a group. Strictly, this galaxy should qualify as a true BGG only when the magnitude gap $\Delta{\cal{M}}_{\rm{2-1}}$ relative to the second-ranked member ($m_2$) exceeds a given threshold.}. This classification highlights how variations in the initial conditions give rise to divergent group assembly pathways.

Groups classified as single-BGG systems are those with a magnitude gap between the first- and second-ranked galaxies, $\Delta{\cal{M}}_{2-1}$, greater or equal to $0.75$ mag at $z_{\rm fin}=0$, comprising approximately $38\%$ of the total sample. Double-BGG systems, which account for $16\%$ of our systems, are identified by $\Delta{\cal{M}}_{2-1} < 0.5$ mag and a magnitude gap between the second- and third-ranked galaxies $\Delta{\cal{M}}_{3-2} \geq 0.75$ mag, indicating the presence of two similarly bright galaxies that clearly stand out from the rest of group members. The remaining $46\%$ of the groups are categorized as non-BGG systems, typically exhibiting multiple ongoing mergers and lacking truly dominant galaxies, as both $\Delta{\cal{M}}_{2-1}$ and $\Delta{\cal{M}}_{3-2}$ are below $0.75$ mag. All magnitude gaps are computed by converting stellar masses ($m_i$) into $K$-band magnitudes, assuming a constant mass-to-light ratio for galaxies in this band ($\Upsilon_{K,\odot}$) of unity.

Figure~\ref{fig:init_mgal_dist} shows the stellar mass distributions of the four most massive galaxies at the start of the simulations, grouped according to the final classification of their host groups. A two-sample Kolmogorov–Smirnov (KS) test reveals no statistically significant differences between group types for any galaxy rank, except for a marginally significant $p$-value  of $0.05$ between single-BGG and non-BGG groups regarding the $m_1 (z_{\rm ini})$ distributions. The lack of significant bias in the initial conditions is further supported by Fig.~\ref{fig:m1_mtot}, which shows the total group mass as a function of the stellar mass of the first-ranked galaxy at $z_{\rm ini}$ in logarithmic scale. The scatter plot is divided into four quadrants using the median values of both variables, facilitating the visual assessment of the lack of noticeable clustering or segregation among the different group classes.

\begin{figure*}[ht!]
\centering
 \includegraphics[width=17.5cm, height=8.75cm]{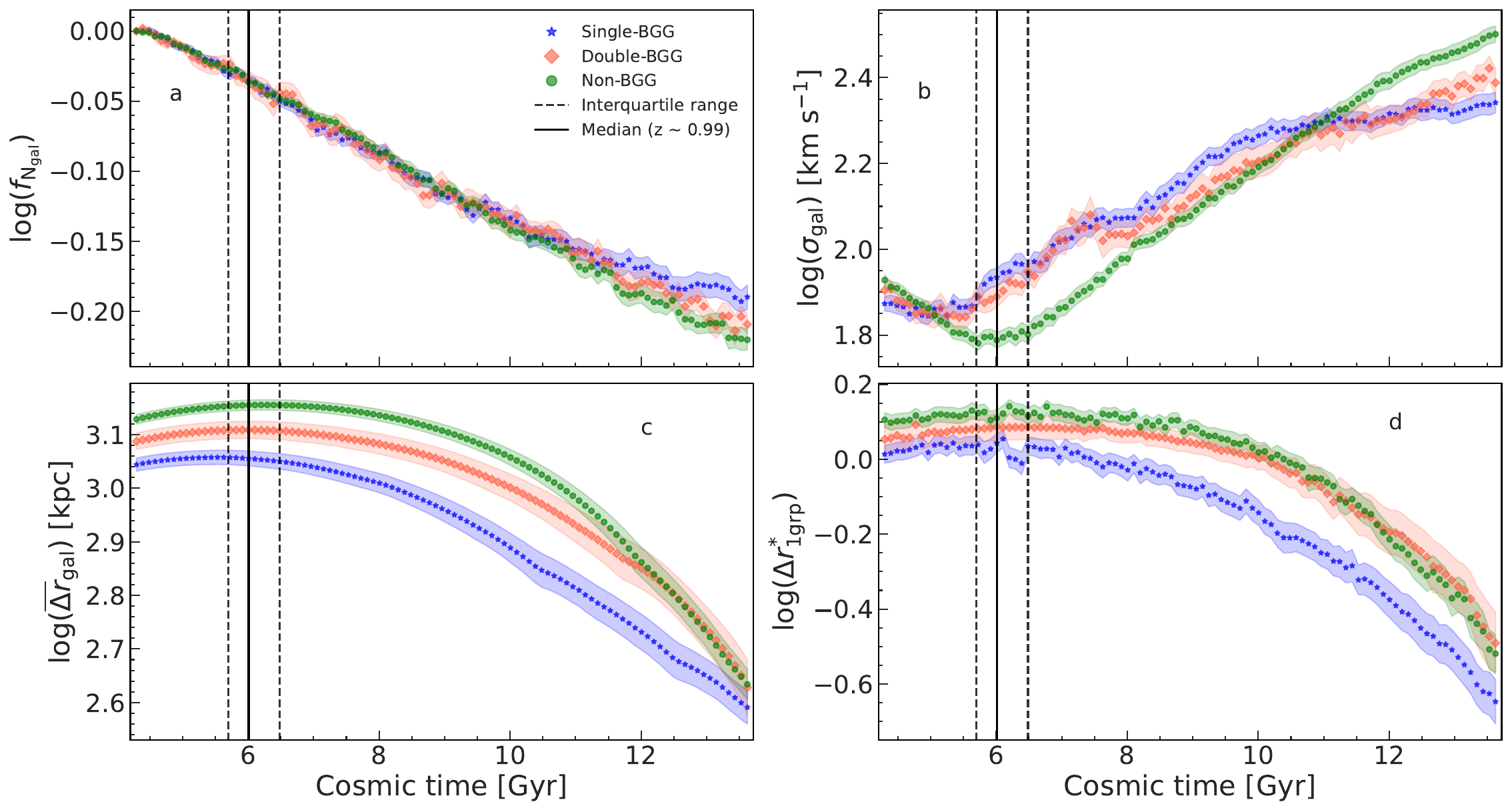}
    \caption{Evolution in the last $\sim 9.3$ Gyr of the runs of various indicators of dynamical state, shown separately for each group class: (a) the fraction of surviving group galaxies relative to $N_{\rm gal}(z_{\rm ref})$; (b) the mass-weighted velocity dispersion of group galaxies; (c) the mass-weighted mean pairwise spatial separation among group members; and (d) the offset between the first-ranked galaxy and the group’s centre of mass, expressed in units of the expected group virial radius (Eq.~(\ref{rvir})). Data points represent the mean values for each class, with the associated transparent shaded bands surrounding them indicating the standard error. The vertical solid and dashed lines in the panels mark, respectively, the $Q_2$, $Q_1$ and $Q_3$ quartiles of the distribution of turnaround times. Group classes are colour-coded as follows: single-BGG systems (blue), double-BGG systems (orange), and non-BGG systems (green).}
    \label{fig:grp_prop}
\end{figure*}

\section{Evolution of galaxy group dynamics since $z = 1.5$}
\label{sec:non-photometric_properties}

In this section, we examine the evolutionary histories of several indicators of the dynamical state of galaxy groups, as listed below.

Fraction of surviving group galaxies relative to the number of member galaxies at a reference epoch, $t_{\rm ref}\equiv t(z_{\rm{ref}})$:
\begin{equation}\label{fNgal}
f_{\rm{N_{\rm gal}}}(t)= \frac{N_{\rm gal}(t)}{N_{\rm{gal}}(t_{\rm ref})} \;.  
\end{equation}

Mass-weighted\footnote{\label{fn:mass-weight}Mass weighting better reflects the true gravitational centre of unrelaxed systems and how galaxies interact.} velocity dispersion of group galaxies:
\begin{equation}\label{sigma_grp}
\sigma_{\rm{gal}} (t)= \sqrt{\frac{1}{M_{\rm gal}} \sum_{i=1}^{N_{\rm gal}} m_{i}(v_{i} -\bar{v}_{\rm{gal}})^2}\;,   
\end{equation}
where $v_{i}$ is the 3D velocity of the $i$-th-ranked galaxy\footnote{Galaxy ranks may change over time.} with stellar mass $m_{i}$, $\bar{v}_{\rm{gal}}$ is the mass-weighted mean velocity of all group galaxies, and $M_{\rm{gal}}=\sum_i m_i$ is the total stellar mass in galaxies, all measured at time $t$. 

Mass-weighted\textsuperscript{\ref{fn:mass-weight}} mean pairwise spatial separation among group galaxies: 
\begin{equation}\label{r_gal}
\overline{\Delta r}_{\rm{gal}} (t)=\frac{2}{{M^2_{\rm gal}}}\sum_{i<j}^{N_{\rm gal}-1} \sum_{j=i+1}^{N_{\rm gal}} {m_{i}m_{j}}{r_{ij}}\;,     
\end{equation}
where $r_{ij}$ is the 3D separation between the $i$-th- and $j$-th-ranked galaxies at time $t$.

Offset between the first-ranked galaxy and the group's centre of mass standardized in units of the radius $R_{\rm vir}$ of a virialised group of the same mass at $z_{\rm fin}=0$:
\begin{equation}\label{offset_norm}
\Delta r^*_{\rm 1grp}(t)=\frac{\Delta r_{\rm 1grp}(t)}{R_{\rm vir}}\;,     
\end{equation}
with
\begin{equation}\label{offset}
\Delta r_{\rm 1grp}(t)=\sqrt{{(x_{1} - x_{\rm grp})^2} + {(y_{1} - y_{\rm grp})^2} + {(z_{1} - z_{\rm grp})^2}}\;,     
\end{equation}
the spatial separation between the position of $m_1$ and the group's centre of mass, evaluated at time $t$, and
\begin{equation}\label{rvir}
R_{\rm vir}\equiv R^{\rm su}\left[\frac{M_{\rm grp}}{M^{\rm su}}\right]^{1/3}\;,
\end{equation}
where $R^{\rm{su}}\simeq288.7\,\mbox{kpc}$ is the scaling between simulation units and physical length for the adopted cosmology \citetext{see also Eq.~(7) in \citealt{2016MNRAS.461..321S}}. 

Figure~\ref{fig:grp_prop} illustrates the secular evolution of the aforementioned indicators for each of the three group classes. Subplot~\ref{fig:grp_prop}a (top-left panel) shows that the change in the fraction of surviving group galaxies, $f_{\rm{N_{\rm gal}}}(t)$, is remarkably similar across all group types over the course of the simulations. The only notable deviation occurs in single-BGG systems, which exhibit a slightly more gradual decline in the number of group members beginning around $8.5$ Gyr after the start of the simulations, or equivalently, at a cosmic time of $t\sim 10.5$ Gyr. 

Equation~(\ref{fNgal}) provides a means to estimate the instantaneous merger rate in the groups, defined as the fraction of galaxies involved in mergers in a $dt$, by taking the time derivative
\begin{equation}\label{dotfNmer}
\dot{f}_{\rm Nmer}= 2\frac{N_{\rm gal}(t) - N_{\rm gal}(t+dt)}{N_{\rm gal}(t)} \cdot \frac{1}{dt}\propto\frac{d\log(f_{\rm N_{\rm gal}}(t))}{dt}\;, 
\end{equation}
with the proportionality factor equal to $-2\ln(10)$. Therefore, the approximately linear evolution with cosmic time observed in the logarithm of $N_{\rm gal}(t)$ indicates that the merger activity associated with the initial gravitational collapse of groups may be effectively modelled as a stochastic Poisson process. As such, it is characterized by three key statistical properties: a constant intensity, indicating a steady merger rate over time; memorylessness, meaning that the likelihood of a merger is independent of previous merger events; and independent increments, such that the number of mergers in disjoint time intervals are statistically independent. Among the three group classes, only the single-BGG systems show a mild deviation from the aforementioned trend, in the form of a modest decline in merger activity toward the end of the initial collapse phase. It is important to note that in this and all subsequent plots presenting the evolutionary histories of dynamical indicators, we focus on the redshift interval $(z_{\rm{ref}},z_{\rm fin})=(1.5,0)$, which encompasses the last $\sim 9.3$ Gyr of the runs. This choice excludes the initial $\sim 2.2\,\mbox{Gyr}$ of the simulations, during which our galaxy groups are still expanding with the Hubble flow before reaching turnaround, so gravitational interactions remain negligible. The median turnaround redshift for our sample of simulated groups is $\bar{z}_{\rm{ta}}\simeq 0.99$\footnote{Obtained by minimising the mass-weighted mean separation (Eq.~(\ref{r_gal})). The value of $\bar{z}_{\rm{ta}}\simeq 0.85$ quoted in \citetalias{paperI} was instead determined from the median of the intergalactic separations. This methodological difference has no impact on the results presented in either paper.}, which in the adopted cosmology is equivalent to a cosmic age of $\bar{t}_{\rm{ta}}\simeq 6.0$ Gyr.

Subplot~\ref{fig:grp_prop}b (top-right panel) displays the time evolution of the velocity dispersion of group galaxies, $\sigma_{\rm{gal}} (t)$. In this case, the outcomes reveal more pronounced differences among group classes, with the average trend for non-BGG groups showing the most significant deviations. These differences become noticeable around the turnaround epoch, which roughly coincides with the point at which the internal velocity dispersion of forming galaxy aggregations reaches a minimum. In single- and double-BGG groups, this minimum occurs approximately 1 Gyr earlier than in their non-BGG counterparts and is about $25\%$ higher, suggesting that in these systems virial heating begins to dominate over the coherence of infall motions before turnaround. In contrast, the late turnaround of non-BGG groups is followed by a faster and more sustained rise in velocity dispersion, so that by the end of the simulations, they exhibit the highest values of the three group classes. This indicates that the interactions, mergers, and tidal shocks occurring among member galaxies during the early pre-virialisation phase of these systems generate a more chaotic internal velocity field, which in turn delays the emergence of a dominant central object.

The bottom panels of Fig.~\ref{fig:grp_prop} show the time evolution of two metrics that assess the dynamical state of the groups based on the relative positions of their member galaxies. Panel~\ref{fig:grp_prop}c (bottom-left) presents the evolution of the mean intergalactic spatial separation, $\overline{\Delta r}_{\rm{gal}}(t)$. We can see that even in the earliest epochs, the three group classes exhibit noticeable differences in the average values of this parameter, with single-BGG groups showing the smallest pairwise separations and non-BGG groups the largest. This circumstance simply reflects the initial conditions of our simulations. Specifically, that the algorithm used to assign the initial number of galaxies to each group (see Sect.~\ref{ssec:models}) accounts for the fact that, above halo masses $\sim 10^{12}\,$\msun, more massive systems are less efficient at cooling baryons and forming galaxies \citep[see e.g.][]{Papastergis_2012,2013MNRAS.428.3121M}. As a result, single-BGG groups -- which tend to be less massive (see Fig.~5 in \citetalias{paperI}) -- are those created with a higher number density of member galaxies, whereas the opposite holds for non-BGG groups. 

Fig.~\ref{fig:grp_prop}c also shows that shortly after turnaround, the mean intergalactic separation in all group classes enters a phase of progressively steeper decline that persists for the remainder of the simulations. This decline is initially more pronounced in single-BGG groups, a consequence of having a single dominant gravitational centre that drives a more efficient collapse. In contrast, double-BGG and non-BGG systems, which contain more than one competing centre of gravitational attraction, exhibit most of the time, a more gradual decrease in the average intergalactic separation as galaxies are pulled in different directions. Nonetheless, by the end of the simulations, the values of this metric converge rapidly across all group classes, consistent with the intended setup of our experiments, where all groups are configured to collapse to a point by the final snapshot.

\begin{figure*}
\centering
 \includegraphics[width=17.5cm, height=8.75cm]{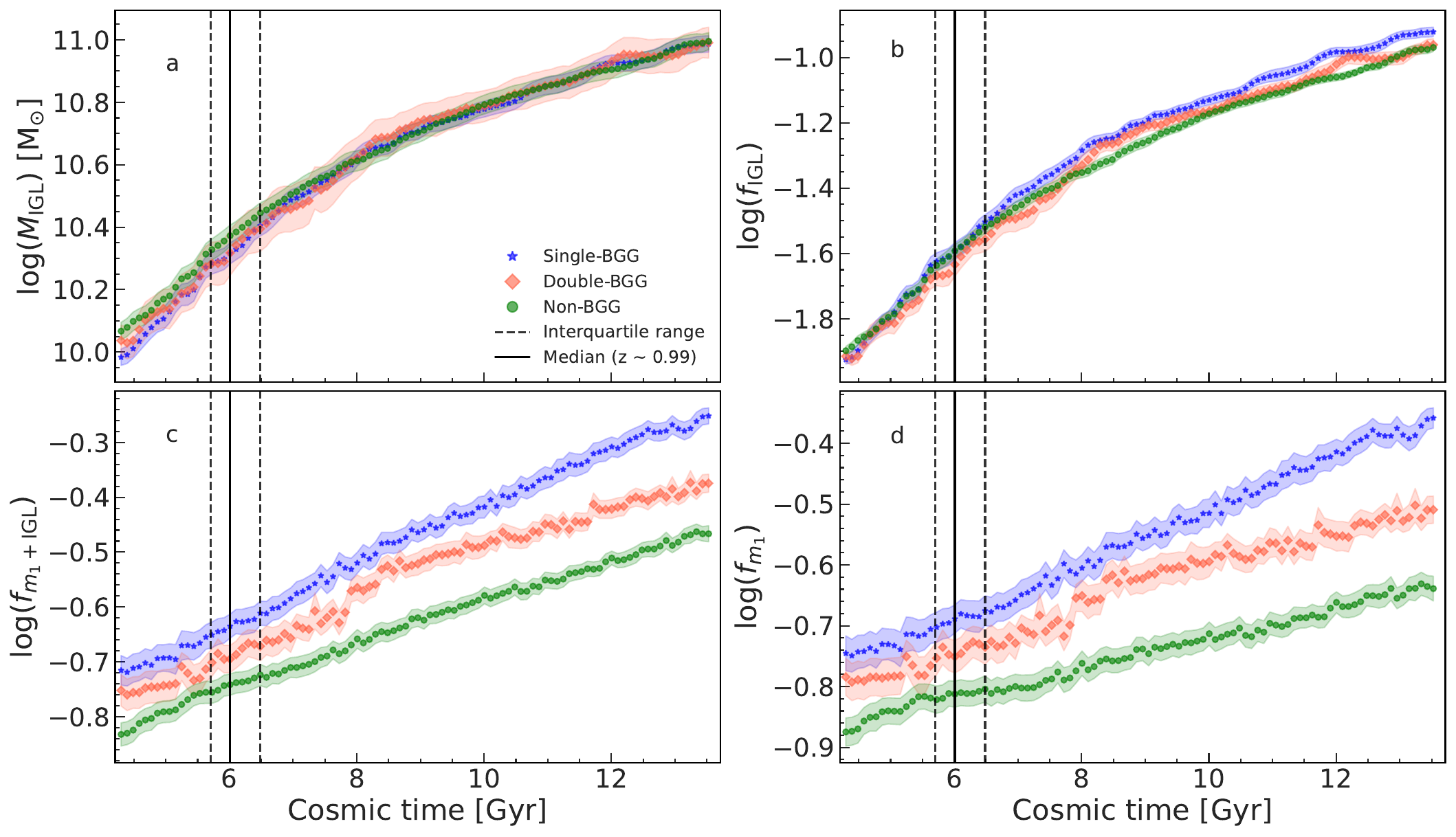}
    \caption{Temporal evolution of various dynamical state indicators based on the IGL, shown separately for each group class: (a) the total IGL mass; (b) the IGL mass fraction; (c) the $m_1$ plus total IGL mass fraction; and (d) the $m_1$ mass fraction. All luminous mass fractions are expressed relative to the invariant total stellar mass of the groups. Data points represent the mean values for each class, with the associated transparent shaded bands surrounding them indicating the standard error. The central vertical solid and dashed left and right lines in the panels mark, respectively, the $Q_2$, $Q_1$ and $Q_3$ quartiles of the distribution of turnaround times for the entire group sample. Group classes are colour-coded as in Fig.~\ref{fig:grp_prop}.}
    \label{fig:igl_prop}
\end{figure*}

For its part, panel~\ref{fig:grp_prop}d (bottom-right) displays the spatial offset between each group's centre of mass and the location of its most significant galaxy, $\Delta r^*_{\rm 1grp}(t)$, calculated using Eqs.~(\ref{offset_norm})--(\ref{rvir}). The overall evolution of the average values of this parameter closely parallels that of the mean intergalactic separation, with the trajectories again depending on group class. In this case, however, the differences between double- and non-BGG groups are modest, with their evolutionary trends remaining statistically consistent throughout the simulation. As shown in this figure, the average spatial offset in both classes keeps rising slightly until at least $\sim 3$ Gyr after turnaround, when it begins to decrease appreciably. Single-BGG groups, in contrast, which display the smallest average displacements throughout the entire time sequence, follow immediately after turnaround a steeper decline indicative of a more rapid progression toward dynamical relaxation. By the end of the runs, the mean central offset in this group class declines to $\sim 25\%$ of the virial radius expected for a system of the same mass, whereas non-BGG and double-BGG groups retain somewhat larger residual values of about $0.4\,R_{\rm vir}$. This difference stems primarily from the weakening of the link between $m_1$ and the group’s centre of mass in these latter group classes, owing to the presence of other galaxies of comparable gravitational influence.

Overall, the evolutionary patterns presented in Fig.~\ref{fig:grp_prop} further reinforce the conclusion, previously put forward in \citetalias{paperI}, that there exists a clear correlation between our classification of actively merging groups and their dynamical state. Specifically, single-BGG systems tend to progress more rapidly towards dynamical relaxation, whereas double-BGG groups and, especially, non-BGG groups undergo a more gradual and less efficient evolution.

In the next two sections, we examine the long-term behaviour of other well-known estimators of hierarchical group assembly, focusing specifically on those related to the formation of the IGL and the emergence of the most massive galaxy within the groups.

\section{Evolution of IGL since $z = 1.5$}
\label{sec:photometric_properties}

\subsection{Secular evolution of IGL-based dynamical indicators} \label{ssec:dynamical_indicators}

\begin{figure*}
\centering
 \includegraphics[width=18cm, height=5.0cm]{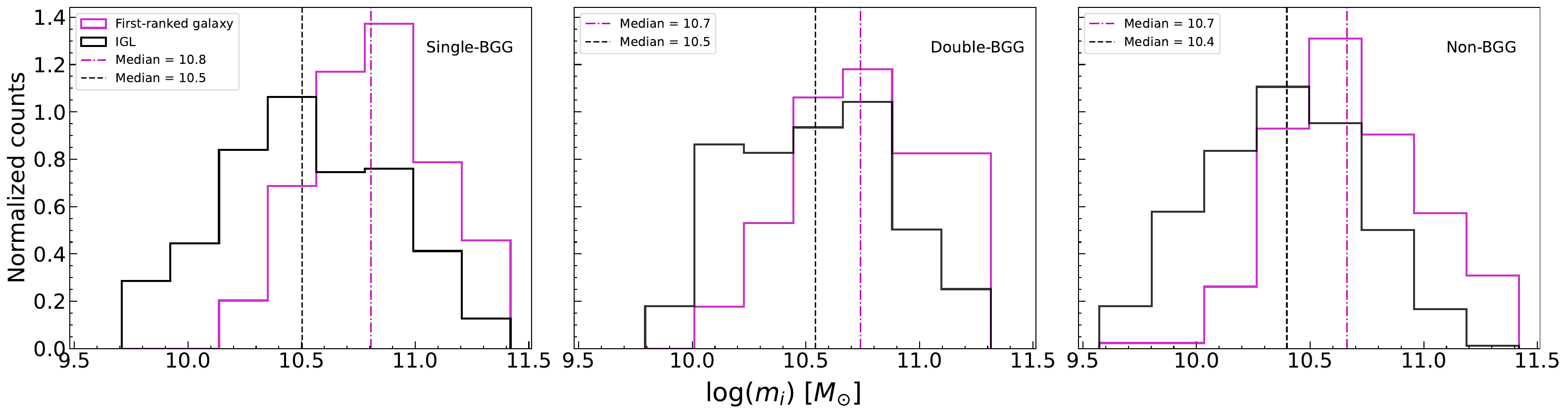}
    \caption{Histograms showing the luminous mass distribution of precursor galaxies at $z_{\rm ini}=3$ that contribute at least $5\%$ to the final mass of either the first-ranked group galaxy $m_1$ (solid magenta lines) or the IGL (solid black lines), with median values indicated by vertical dot-dashed and dashed lines, respectively. The distributions are shown separately for each group class: single-BGG (left), double-BGG (middle), and non-BGG (right) panels.}
    \label{fig:m1_igl_cont_gal}
\end{figure*}

In this section, we investigate the secular evolution of luminous tracers that offer a direct insight into the dynamical histories of galaxy groups. Chief among these is the total ex situ IGL mass, $M_{\rm IGL}(t)$, and, in particular, its normalized fraction relative to the total stellar mass of the group, $M_\star$:
\begin{equation}\label{figl}
f_{\rm IGL}(t)=\frac{M_{\rm IGL}(t)}{M_{\rm \star}}\;,
\end{equation}  
that we have computed at each time step by averaging the results obtained from three independent orthogonal projection angles (see Sect.~\ref{ssec:identification}). Figures~\ref{fig:igl_prop}a--b present the evolution of the IGL mass and its fractional contribution to $M_\star$ over cosmic time. Both panels show that the average amount of IGL and its growth rate are largely insensitive to group class and total mass. The only notable deviation is a slight divergence in the IGL fractions of single-BGG systems, which tend to exhibit marginally higher values over time. These plots also reveal that the accumulation of IGL, while occurring across all redshifts, proceeds in two distinct phases, being faster in relative terms up to the first half of the analysed time interval, followed by a more subdued but ultimately dominant accumulation over the final $\sim 5$ Gyr of the runs. This latter regime, in good agreement with the results from the cosmological hydrodynamical simulations of \citet{2007MNRAS.377....2M} and semi-analytical model predictions of \citet{Contini+2014}, corresponds to the advanced stages of hierarchical group assembly, when the post-turnaround contraction of intergalactic distances increases the frequency and strength of gravitational encounters conductive to the unbinding of the stars. At the end of the runs the distribution of the total ex situ IGL mass generated spans an interquartile range of $(Q_1,Q_2,Q_3)=(7,10,13)\,\times 10^{10}\,$\msun, corresponding to IGL fractions of $\sim 6$--$18\%$, and a median of $11.3\%$. These values are broadly consistent with those reported in previous studies of systems with comparable mass and dynamical state \citep{2005MNRAS.358..949Z, 2007MNRAS.377....2M, 2011ApJ...732...48R, 2015MNRAS.449.2353B, 2016MNRAS.461..321S, 2022FrASS...972283A, Ahad+2025}, reinforcing the view that our simulations reproduce realistic levels of diffuse light. A more detailed comparison of these results with both observational and theoretical estimates of diffuse-light fractions in galaxy systems is provided in \citetalias{paperI}.
 
Another informative tracer of the impact of mergers and tidal interactions on the redistribution of luminous material in bound galaxy systems is the combined stellar mass fraction locked in the first-ranked galaxy and the diffuse IGL. This metric is widely used in observational studies of large associations, such as galaxy clusters \citep[see e.g.][and references therein]{2024ApJ...965..145Y}, where,  in addition to the well-known challenges posed by disentangling diffuse stellar emission from galaxy outskirts, the large spatial extent of these systems typically restricts measurements to the densest central regions around the first-ranked object. In such cases, it is customary to treat the accumulation of stellar mass, in both  $m_1$ and its surrounding diffuse envelope, as a single, combined component to probe the formation history of these systems. As shown in Fig.~\ref{fig:igl_prop}c, the evolution of the $m_1+\mbox{IGL}$ mass fraction in our simulations follows an approximately linear growth in logarithmic scale, though with notable variations across group classes. These differences are in the form of a progressive increase in both the normalization and slope of the trends from non-BGG groups to double-BGG systems, and ultimately to single-BGG groups. We caution, however, that in systems with relatively modest IGL content -- such as the simulated groups studied here -- this metric may largely reflect the assembly history of the first-ranked galaxy, providing limited insight into the independent evolution of the diffuse light component, even when its total mass is included. This is reflected in the close similarity between the evolutionary trends reported in panels c and d of Fig.~\ref{fig:igl_prop}  \citetext{see also subplots (4) and (5) in Fig.~4 of \citealt{2024ApJ...965..145Y}}. 

\subsection{The distinct growth histories of the IGL and $m_1$} \label{ssec:growth_histories}

It is important to recall that, despite their intertwined origins, the first-ranked galaxy and the diffuse stellar component follow distinct growth histories, involving different populations of progenitors, as demonstrated in \citetalias{paperI} \citep[see also, for instance,][]{2018MNRAS.479..932C}. This distinction is illustrated in   Fig.~\ref{fig:m1_igl_cont_gal}, which compares the luminous mass distributions of the progenitor galaxies at the beginning of the simulations that contribute at least $5\%$ to the final mass of either component, shown separately for each group class. As the figure reveals, the relevant progenitors involved in the formation of the main galaxy and the IGL differ significantly in both their characteristic (median and modal) masses and their overall mass ranges, the latter being notably broader and skewed towards lower values for the diffuse component. A two-sample KS test supports this visual impression, yielding $p$-values $\leq 0.001$ in all cases.

\begin{figure}
\centering
 \includegraphics[width=9cm, height=6.88cm]{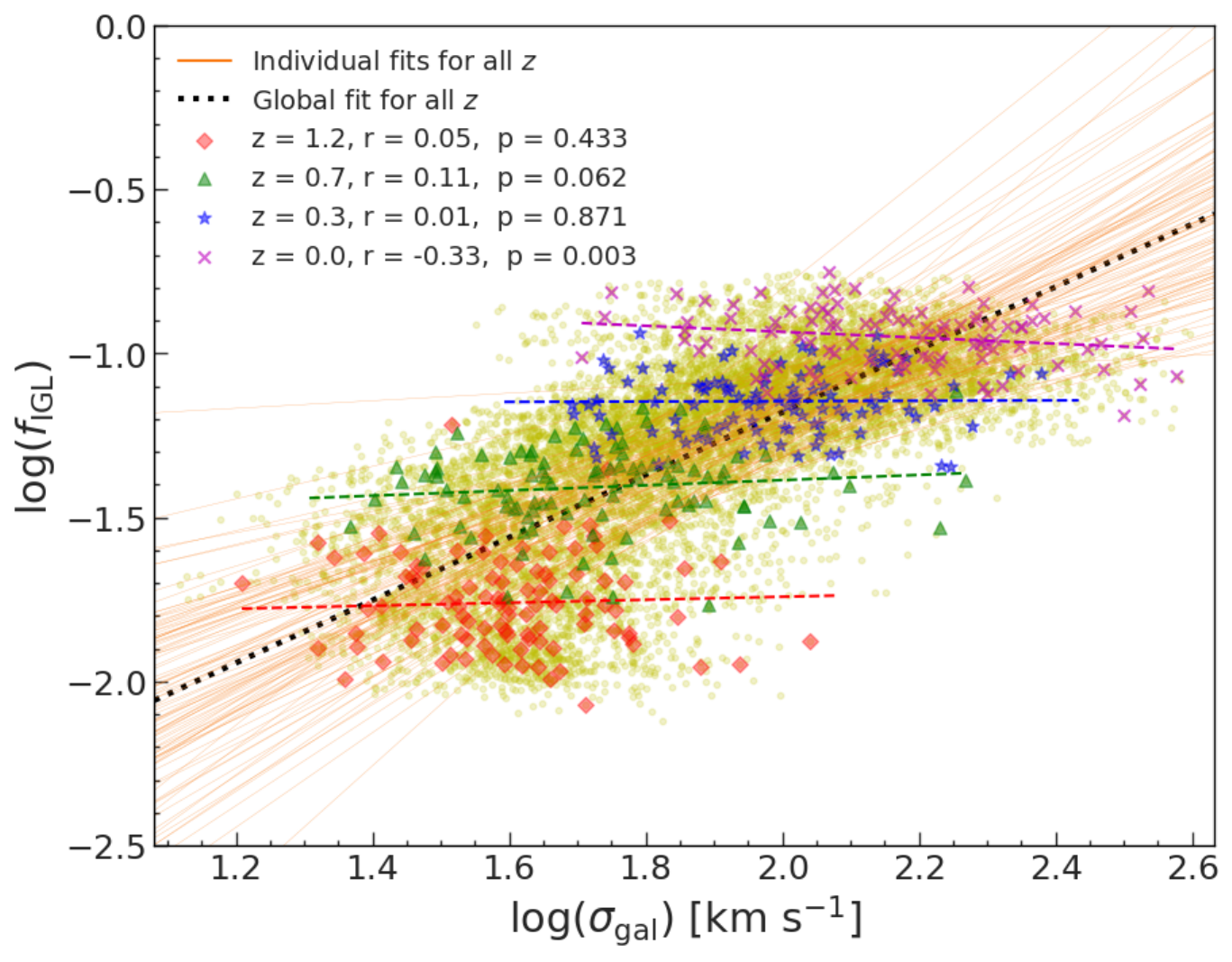}
    \caption{Temporal evolution of the relationship between the IGL mass fraction, $f_{\rm IGL}(t)$, and the mass-weighted velocity dispersion of group galaxies, $\sigma_{\rm{gal}}(t)$. Thin solid orange lines represent linear fits to the secular evolution of these quantities in individual groups, while the dotted black line shows the global trend. Dashed lines indicate the best linear fits of this relationship at four specific redshifts: $z = 1.2$ (red, diamonds), $z = 0.7$ (green, triangles), $z = 0.3$ (blue, stars), and $z = 0$ (purple, crosses), with line colours matching the corresponding data symbols. Each cosmic epoch-specific fit is annotated with its Pearson $r$ correlation coefficient and $p$-value. For clarity, only data corresponding to the projection onto the $XY$ plane are displayed, although the reported fit statistics take into account all three Cartesian projections.}
    \label{fig:vdisp_fIGL}
\end{figure}

\begin{figure}
\centering
 \includegraphics[width=8.8cm, height=14.3cm]{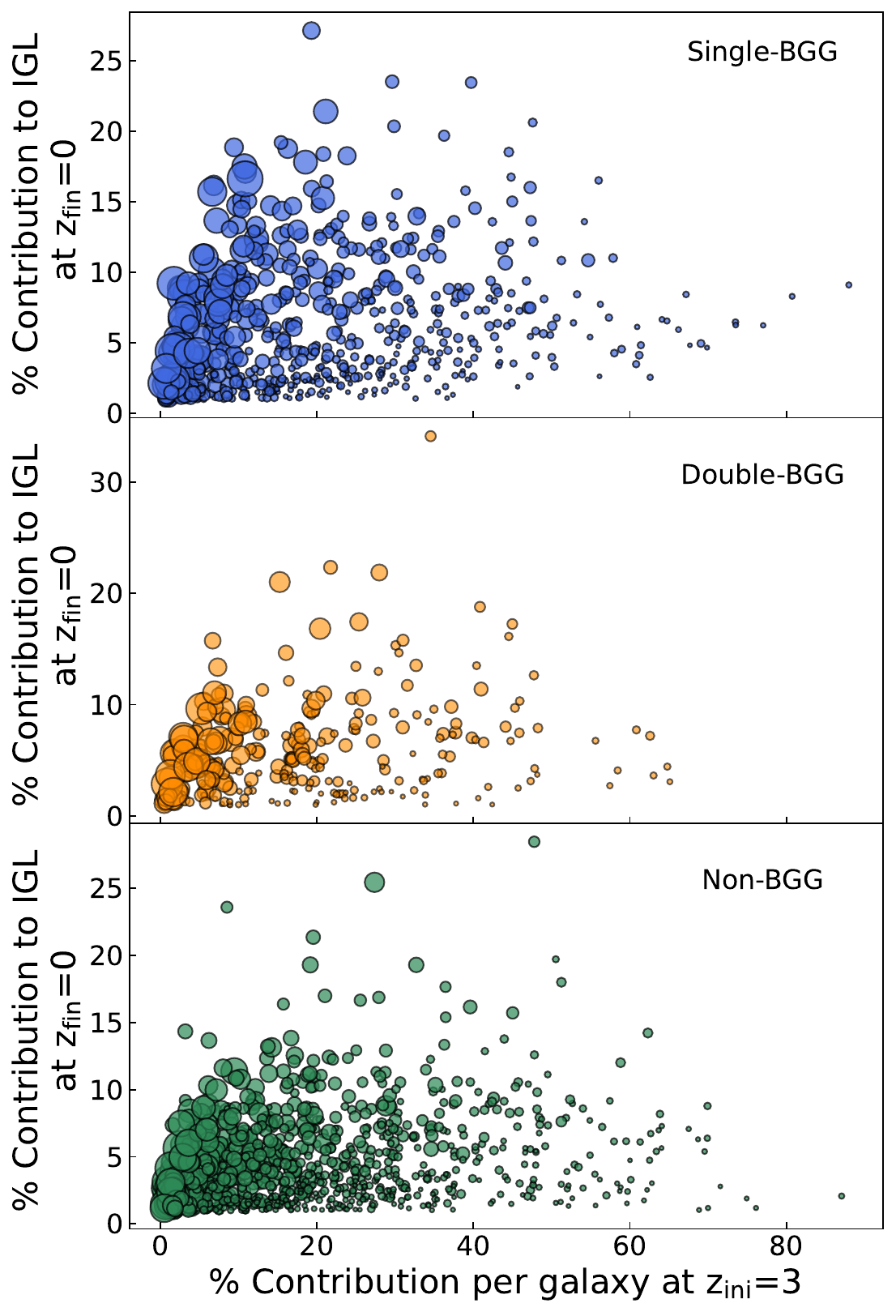}
    \caption{The fraction of the stellar mass in each galaxy at $z_{\rm ini} = 3$ that contributes to the IGL mass at $z_{\rm fin} = 0$ vs the fraction of the final IGL mass that this contribution represents, categorized by group class. From top to bottom, panels correspond to single-BGG groups, double-BGG groups, and non-BGG groups. Point sizes scale with the stellar mass of the contributing galaxies at $z_{\rm ini}$, while point colours follow the same scheme used throughout the paper.} 
    \label{fig:mass_cont}
\end{figure}

Another result reported in \citetalias{paperI} was the confirmation that IGL growth is promoted by low-velocity encounters between member galaxies, which enhance the efficiency of gravitational interactions  \citep[see also e.g.][]{Mihos:2003,BinneyTremaine2008,2016ApJS..225...23S}. This conclusion was supported by the detection of a modest but highly significant negative linear correlation between the logarithms of the IGL mass fraction and the mass-weighted velocity dispersion of group galaxies at the end of the simulations (Pearson coefficient $r = -0.33$ and $p$-value $0.003$). Here, we expand on this result with Fig.~\ref{fig:vdisp_fIGL}, which displays the temporal evolution of the $\log(f_{\rm IGL}(t))$–$\log(\sigma_{\rm gal}(t))$ relation for both individual groups and the full sample. In this figure, linear fits to the data at four representative cosmic epochs ($z = 1.2$, $0.7$, $0.3$, and $0$) trace the emergence of the correlation over time. These epoch-specific fits and the accompanying statistics demonstrate that the two dynamical metrics remain largely uncorrelated until $z\simeq 0.3$, implying that the negative correlation observed at $z=0$ necessarily develops during the last $\sim 3.5$ Gyr of the simulations, coinciding with the main phase of IGL assembly (see Fig.~\ref{fig:igl_prop}a). In this regard, we speculate that our simulations of isolated groups might underestimate the overall merger rate -- and thus the growth of both the IGL fraction and $m_1$ -- compared to fully cosmological simulations, where groups are mostly assembled within filaments of the cosmic web. Cosmic filaments channel galaxies along preferred directions, raising local number densities and encounter cross-sections near the filament spine. This coherent streaming keeps relative velocities more 1D and typically lower than in more isotropic large-scale flows, enhancing gravitational capture. Both effects are known to boost close-pair and merger frequencies and contribute to increasing the range of galaxy sizes within groups \citep{Vijayaraghavan+2013,Liu+2024}.

\begin{table*}
\centering
     \caption[]{Galaxies at $z_{\rm ini}=3$ contributing at least $5\%$ of the total amount of IGL at $z_{\rm fin}=0$, categorized by group class.}
        \label{tab:major_cont}
        \begin{tabular}{l|c|c|c|c}
                \toprule
    \bf Group class & \makecell{\bf No. of contributing \\\bf galaxies} & \makecell{\bf Mass of contributing \\\bf galaxies [$\bf 10^{10}\,$\msun]}  & \makecell{\bf Ranks of contributing \\\bf galaxies} & \makecell{\bf Mass  contribution \\\bf per galaxy [\%]}\\
                \midrule
           Single-BGG & $8^{+1}_{-2}$  &   $3.2^{+2.1}_{-1.4}$  & $7^{+5}_{-4}$ & $8.8^{+3.0}_{-2.0}$ \\
           & & &\\
           Double-BGG & $8^{+1}_{-1}$  &   $3.5^{+2.8}_{-1.7}$  & $8^{+5}_{-4}$ &  $8.0^{+2.3}_{-1.3}$       \\
           & & &\\
           Non-BGG & $8^{+0}_{-2}$  &   $2.5^{+1.7}_{-1.1}$  & $11^{+6}_{-5}$ &  $7.4^{+1.8}_{-1.3}$        \\
        \bottomrule
        \noalign{\smallskip}
     \end{tabular}
\end{table*}

\begin{table}
\begin{threeparttable}
\caption[]{Likelihood\tnote{a} of a common mass distribution of progenitor galaxies\tnote{b}.}
	\label{tab:ks_test}
    \centering
    \renewcommand{\arraystretch}{1.6}
    \begin{tabular}{l|c|c|c|c|c}
        \hline
        \multicolumn{1}{c|}{\bf{Group class}} & \multicolumn{5}{c}{\bf{Component}}
        \\\cmidrule{2-6} 
        \multicolumn{1}{c|}{\bf{comparison}} & {\bf IGL} & \multicolumn{1}{c|}{$\mathbf{m_1}$} & \multicolumn{1}{c|}{$\mathbf{m_2}$} & \multicolumn{1}{c|}{$\mathbf{m_3}$}  & \multicolumn{1}{c}{$\mathbf{m_4}$} \\ 
        \hline
        Single- vs double-BGG   & $.66$ & $.29$   & $.04$   & $.29$  & $.08$ \\ 
        Single- vs non-BGG  & $.001$ & $.001$  & $.28$   & $.14$  & $.66$ \\ 
        Double- vs non-BGG  & $.001$ & $.32$  & $.30$   & $.86$  & $.32$ \\ 
        \hline
    \end{tabular}    
    \begin{tablenotes}
        \small 
       \item [a]$p$-values inferred from a two-sample KS test.
       \item [b]Providing $\geq 5\%$ of the final mass of the component.
     \end{tablenotes}
  \end{threeparttable}
\end{table}

To gain more detailed insights into the formation history of the IGL, we cross-matched the stellar particles bound to galaxies in our forming groups at the beginning of the simulations with those identified as part of the IGL at the final time step. This procedure enabled us to track which galaxies were disrupted over the course of the runs and subsequently contributed to the diffuse stellar component. 

Figure~\ref{fig:mass_cont} shows, for each group class separately, the fraction of each galaxy’s stellar mass present at $z_{\rm ini} = 3$ that ends up in the final IGL, plotted against the fraction of the total IGL mass contributed, with point sizes indicating the initial luminous mass of each galaxy. The three scatter plots highlight a general trend: relatively small fractions of stellar mass stripped from the most massive progenitors contribute substantially to the IGL, whereas low-mass counterparts, despite often losing a larger share of their stellar content, usually add less in absolute terms \citep[see also, e.g.][]{2007MNRAS.377....2M, Martel+2012}.

There are, however, differences in IGL growth across the various group types that we summarise in Table~\ref{tab:major_cont}, which lists the characteristics of IGL progenitors at the start of the simulations that provide at least $5\%$ of the final mass of this component (see also Fig.~\ref{fig:m1_igl_cont_gal}). For each group class, the table reports the median and interquartile ranges of the number of significant contributors, along with their stellar masses, initial ranks, and individual fractional contributions to the IGL. The data reveal that, in single- and double-BGG groups, the bulk of the IGL is built predominantly from more massive progenitors, with over half of the main contributors belonging to the ten most massive galaxies at the start of the simulations. Moreover, the fractional contribution per galaxy is notably higher in single-BGG groups, reflecting the stronger gravitational interactions driving their assembly. By contrast, non-BGG groups, while hosting a comparable number of significant progenitors, draw them from a broader mass range; these progenitors tend to be less massive and individually supply smaller fractions of their mass to the IGL. Indeed, in such systems roughly one third of the IGL is assembled from small galaxies, each contributing less than $5\%$ to the total diffuse component. This explains how non-BGG groups can build IGL masses and fractions comparable to those of the other group classes, and highlights the important role that low-mass galaxies can play in the formation of diffuse stellar light in some group environments \citep[e.g.][]{2020A&A...640A.137R}.

These findings for non-BGG groups are consistent with observational studies of diffuse stellar light in nearby dynamically unrelaxed galaxy systems, such as Leo I (around M105) and the Virgo subcluster B (around M49). In these environments, the IGL is dominated by old ($\gtrsim 10$ Gyr), metal-poor stellar populations, suggesting an origin primarily from low-mass satellites with typical stellar masses of a few times $10^8\,$\msun, as inferred from the low-mass end of the mass–metallicity relation \citetext{see \citealt{2022FrASS...972283A} and references therein}. By contrast, our simulations indicate that in single- and double-BGG groups the IGL is generally sourced by more massive progenitors, in line with what is observed in more dynamically evolved systems such as the Hydra I cluster, where the outer envelope of the central galaxy NGC 3311 is believed to be assembled mainly from relatively massive, $\alpha$-enhanced satellites \citep{Barbosa+2021}. Interestingly, this diversity in progenitor masses is also echoed in recent theoretical studies based on high-resolution hydrodynamical simulations. While some works emphasize that the bulk of this component in groups and clusters originates from galaxies with stellar masses around $10^{11}\,$\msun\ \citep[e.g.][]{Brown+2024}, others show that lower-mass objects can also make a non-negligible contribution, with their relative importance varying widely, not only with the total mass of the host system, but also with its assembly history and with the method used to identify the unbound stellar component \citep[e.g.][]{Ahvazi+2024}.

Table~\ref{tab:ks_test} reports the results of KS tests comparing the stellar‐mass distributions of progenitor galaxies that each contribute at least $5\%$ to the IGL across our different group classes. The $p$‑values in the second column demonstrate that the main IGL progenitors in non‑BGG groups are statistically distinct from those in both single‑ and double‑BGG systems, further underscoring the diversity of evolutionary channels through which the IGL is assembled in our simulated groups.

In the following two sections, we analyse the evolution of the stellar mass of top-ranked galaxies within our group sample and assess the effectiveness of the corresponding luminosity gaps as quantitative indicators of the dynamical age of galaxy systems.

\section{Evolution of top-ranked galaxies since $z = 1.5$} \label{sec:top-ranked-galaxies}

\begin{figure}
\centering
 \includegraphics[width=8.5cm, height=17cm]{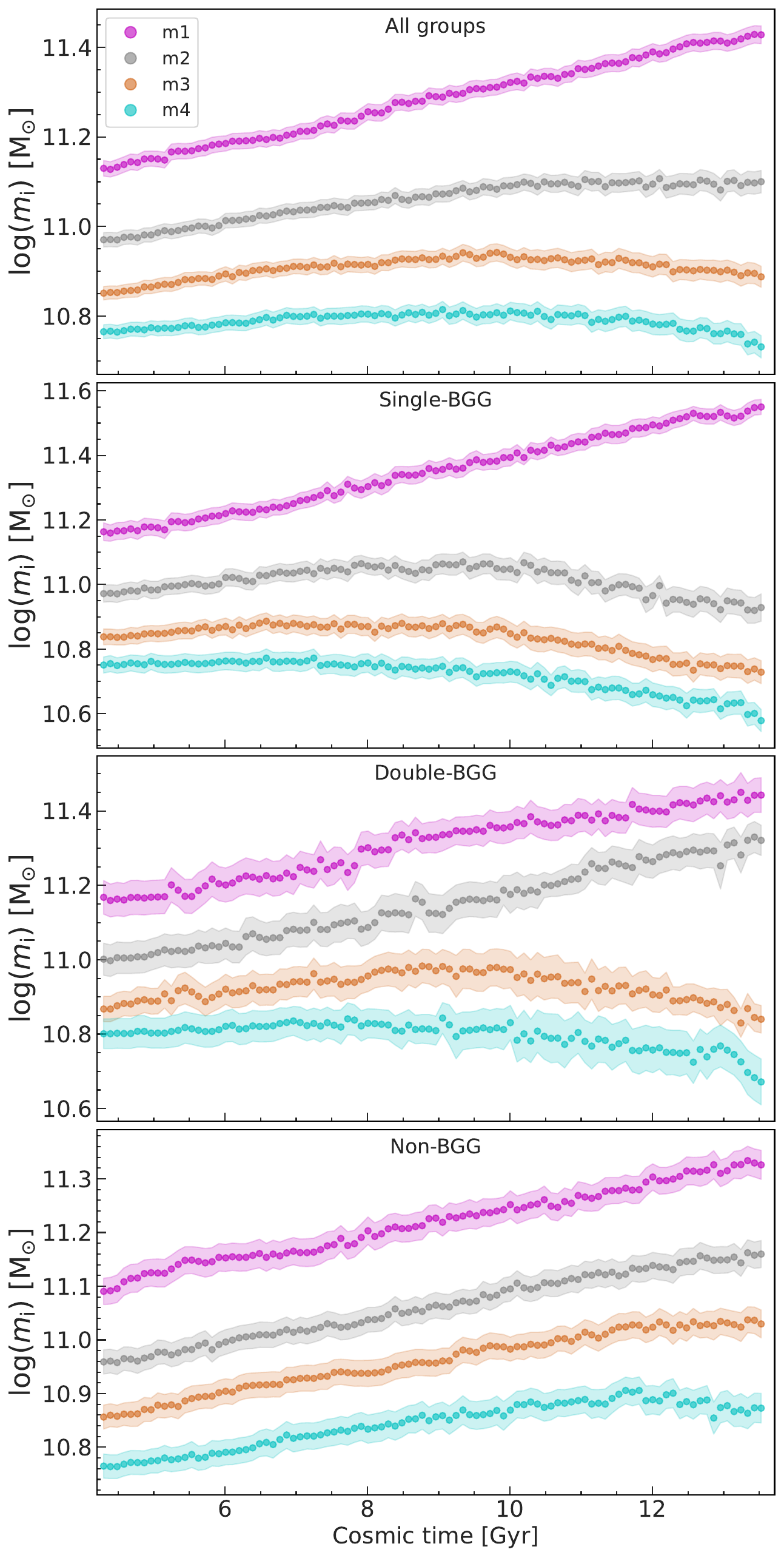}
    \caption{Temporal evolution of the average stellar mass of the four most massive (brightest) galaxies identified in each of our galaxy groups. From top to bottom, panels correspond to all groups combined, single-BGG groups, double-BGG groups, and non-BGG groups. Data points represent the mean stellar mass of $m_1$ (magenta), $m_2$ (grey), $m_3$ (brown), and $m_4$ (cyan) galaxies, with the associated shaded regions indicating the standard error.} 
    \label{fig:mgal_t}
\end{figure}

\begin{figure*}
\centering
 \includegraphics[width=16.8cm, height=10cm]{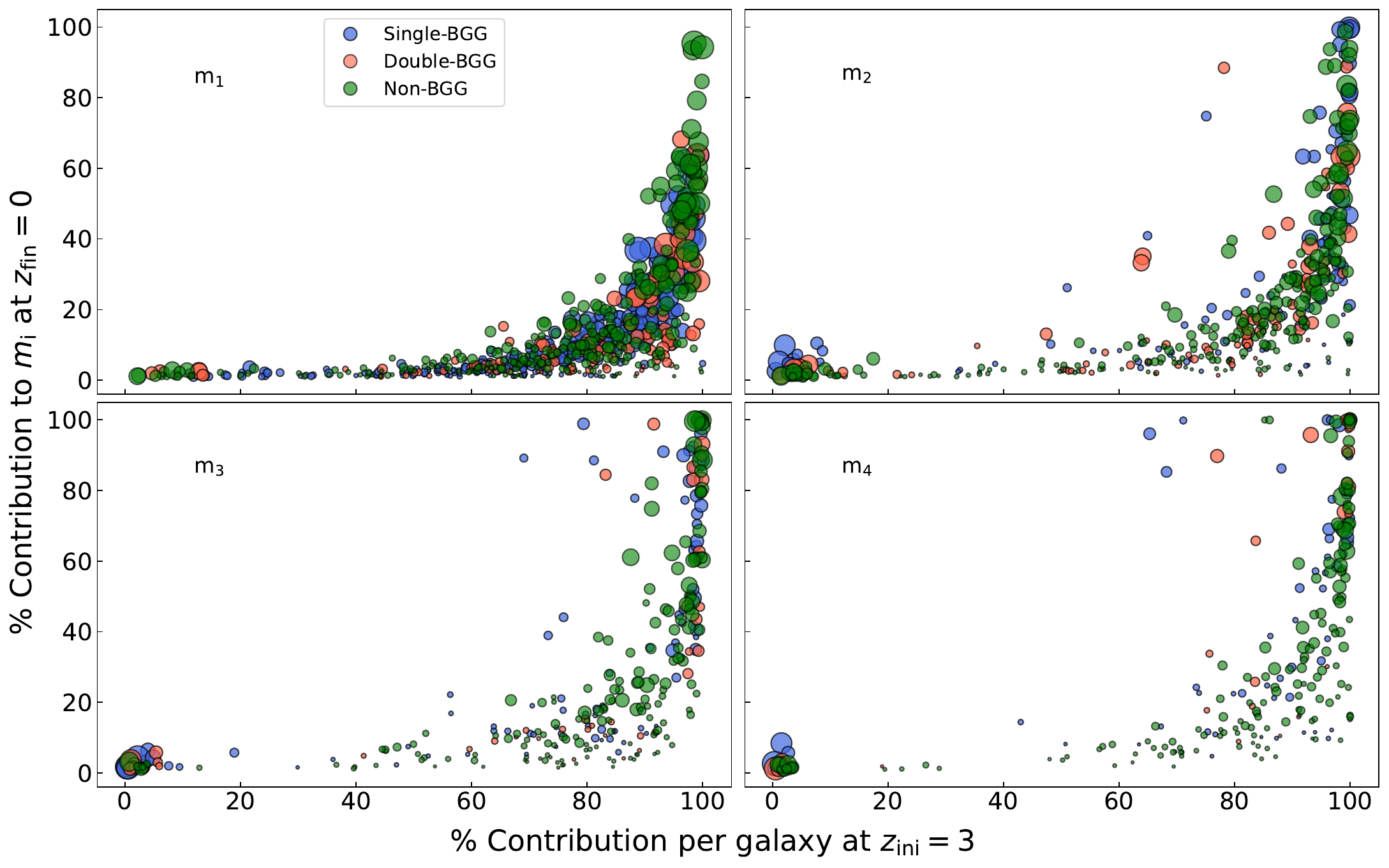}
    \caption{Fractional contribution of individual galaxies at $z_{\rm ini}=3$ to the stellar mass at $z_{\rm fin}=0$ of the 1st- (top-left panel), 2nd- (top-right panel), 3rd- (bottom-left panel), and 4th-ranked galaxy (bottom-right panel), categorized by group class. Point sizes are proportional to the stellar mass at $z_{\rm ini}=3$ of the contributing galaxies. Group classes follow the same colour scheme adopted throughout the paper.}
    \label{fig:mcont_corr}
\end{figure*}

In dynamically complex galaxy associations, the notion that hierarchical growth predominantly affects the most massive (brightest) galaxy while leaving satellites largely unaffected may occasionally apply, but fails to hold as a general rule under closer examination. This limitation is illustrated in Fig.~\ref{fig:mgal_t}, which traces the evolution of the average stellar mass of the four brightest galaxies in each group of our sample. As the different panels show, the dynamical evolution of galaxy groups impacts not only their first-ranked galaxies but also significantly shapes the growth of other high-ranked members, with the magnitude and character of this impact varying markedly across group classes. Consistent with their role as central halo occupants, first-ranked galaxies in all group classes display an approximately linear increase in the logarithm of their stellar mass over cosmic time. The most prominent systematic trend is a gradual decline in growth rate from single-BGG to double-BGG and non-BGG systems (see also Fig.~\ref{fig:igl_prop}d). This broad similarity in the evolution of $m_1$, however, breaks down when considering the most massive satellites. In single-BGG groups, second-ranked galaxies grow steadily until roughly $4$ Gyr before the end of the simulations, after which their masses begin to decline. By contrast, their counterparts in double- and non-BGG groups exhibit uninterrupted mass growth throughout the simulations, with those in double-BGG systems even surpassing the relative growth rate of their respective centrals. A similar pattern of sustained, albeit slower, mass increase is seen for the third- and fourth-ranked members of non-BGG groups, although in the latter this trend reverses toward the end of the simulations, with their average stellar mass declining during the final $\sim 2$ Gyr. In single- and double-BGG systems, by comparison, these satellites typically undergo a short initial growth phase, followed by a more or less extended period of mass decline.

These divergent evolutionary paths of the principal satellites can be attributed to two main evolutionary channels: (a) incorporation into the central or a higher-ranked galaxy via merger, which necessarily entails a rank swap with a less massive companion, or (b) net mass loss through tidal disruption or galaxy harassment that outweighs any merger-driven growth, a process that may or may not also alter the rank. An analysis of the full assembly histories of top-ranked galaxies across all group classes shows that the vast majority grow predominantly through ex situ accretion, with the global fractions of objects that have increased their mass via mergers by $z_{\rm fin}$ ranging from $100\%$ for the brightest members to about $63\%$ for fourth-ranked galaxies, while significant stellar mass loss affects no more than $8\%$ of them (see also below and Sect.~\ref{ssec:mgap_assembly_histories}). These findings are consistent with both theoretical and observational work underscoring the importance of major mergers in BGG/BCG growth \citep[e.g][]{2007MNRAS.375....2D,Groenewald+2017}. Notably, the results reported in Fig.~\ref{fig:mgal_t} agree closely with the recent machine-learning-based study of \citet{Angeloudi+2024}, trained on a cosmological simulation, which reveals that in the nearby Universe the most massive galaxies ($M_\star > 10^{11}\,$\msun) assemble a significant fraction of their mass via mergers. By contrast, substantial tidal disruption is confined mainly to the lower-mass population, consistent with earlier numerical studies \citep[e.g.][]{Martel+2012}.

\begin{figure}
\centering
 \includegraphics[width=9cm, height=6.88cm]{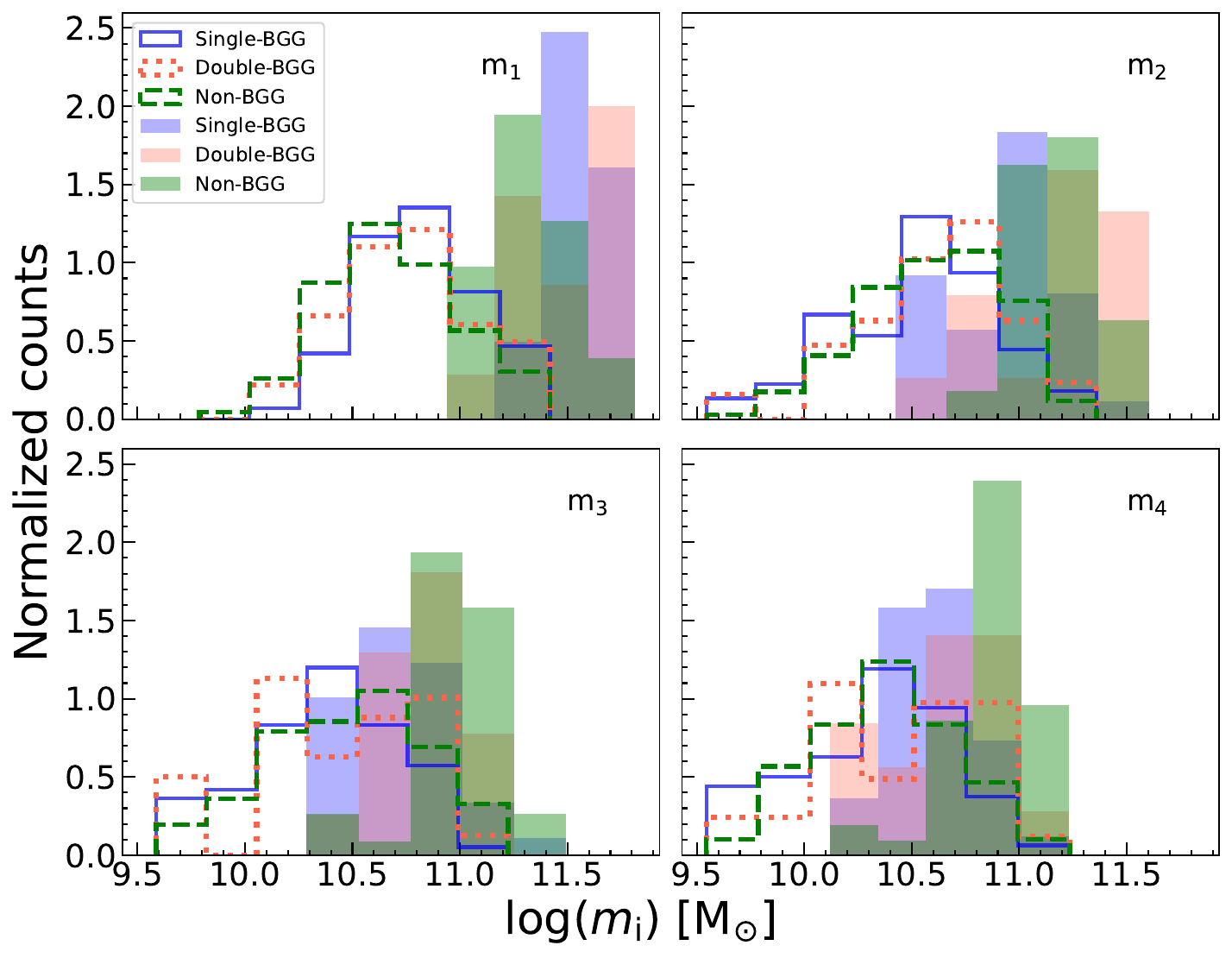}
    \caption{Stellar mass distributions of progenitor galaxies at $z_{\rm ini}=3$ contributing at least $5\%$ to the final stellar mass of the four top-ranked galaxies at $z_{\rm fin}=0$, shown separately by group class. In each panel, transparent-shaded histograms are overlaid to show the stellar mass distributions of the corresponding top-ranked galaxies at the end of the simulations. The initial mass distributions of these galaxies are provided in  Fig.~\ref{fig:init_mgal_dist}. The coding of colours and line styles of group classes is the same as in Fig.~\ref{fig:init_mgal_dist}.}
    \label{fig:mgal_cont_hist}
\end{figure}

The fractional contribution of all progenitors present at the start of the simulations to the stellar mass of each one of the four top-ranked galaxies identified in the final snapshots is presented in Fig.~\ref{fig:mcont_corr} by means of scatter plots. Each data point is colour-coded by group class, and its size encodes the initial stellar mass of the contributing galaxy, thereby facilitating visual comparison of the progenitor populations linked to each top-ranked system member. The plots reveal that, while the most massive galaxies at $z_{\rm ini}$ contribute significantly to the formation of $m_1$, numerous lower-mass progenitors also participate in their assembly, albeit with limited impact on its final mass, as expected. For the second- to fourth-ranked galaxies, a commensurate decrease is observed in the mass range of their main progenitors, with the most massive ones usually being excluded. Nonetheless, these satellites also exhibit clear evidence that mergers with significantly smaller companions contribute to their growth. Additionally, as rank declines, instances gradually emerge in which a galaxy’s fractional mass contribution measured at $z_{\rm ini}$ is smaller than the mass fraction it ultimately accounts for in the final remnant. This behaviour typifies those satellites that, rather than merging, lose part of their stellar content to the IGL through tidal interactions with other group members.

Figure~\ref{fig:mgal_cont_hist} displays histograms of the stellar-mass distributions of progenitor galaxies contributing at least $5\%$ to the final mass of the four top-ranked members, separated by group class. For reference, each panel also includes a secondary set of transparent shaded histograms, representing the stellar-mass distributions of the corresponding top-ranked galaxies at the end of the simulations. As in the IGL analysis, we applied a two-sample KS test to assess whether these distributions differ significantly across group types. The $p$-values reported in the last four columns of Table~\ref{tab:ks_test} indicate that the stellar mass distribution of relevant $m_1$ progenitors in non-BGG groups is statistically inconsistent with that of single-BGG systems, reflecting a relative deficit of high-mass galaxies and a surplus of low-mass ones in the former compared to the latter (see the top-left panel of Fig.~\ref{fig:mgal_cont_hist}). A subtler, yet still statistically significant, difference is also observed between the mass distributions of $m_2$ progenitors in single- and double-BGG groups. In this case, the mismatch appears to be driven by a modest but systematic shift toward higher progenitor masses shown by the groups in the double-BGG class. For all other comparisons, we detect no statistically meaningful differences among group classes. On the other hand, the varying degrees of overlap between the progenitor histograms and the distributions of the corresponding top-ranked galaxies at $z_{\rm fin}=0$ in Fig.~\ref{fig:mgal_cont_hist}, reveal the progressively smaller role of mergers in the assembly of galaxies toward lower ranks.

\section{Magnitude gaps as dynamical state indicators of forming groups} \label{sec:evaluating_mgap}

\subsection{Correlations with the mass fractions in $m_1$ and the IGL} \label{ssec:mgap_correlations}

The findings presented in the previous section highlight that not only the central galaxies of groups, but also their most massive satellites, typically participate -- albeit to a lesser extent -- in the hierarchical assembly of these systems. This naturally leads to whether the magnitude gaps commonly used to gauge the degree of dynamical relaxation of fossil groups retain their diagnostic value when applied to more dynamically complex environments.

To explore this, we analysed the correlations between the set of magnitude gaps that can be defined using the six most luminous galaxies in a group, adopting any of the top three ranked members as reference, and two of the properties that most directly trace the group assembly history: the stellar mass fractions in both $m_1$ and the IGL.

As shown in the top panels of the two leftmost columns of subplots in Fig.~\ref{fig:fIGL_fBGG_mgap} (Appendix~\ref{app:A}), among the correlations between $\log(f_{m_1})$ and the magnitude gaps defined with respect to $m_1$ as reference, the one involving $\Delta{\cal{M}}_{\rm{2-1}}$ stands out as relatively strong. This correlation, however, is driven almost entirely by single-BGG groups, which appear clearly segregated from the rest\footnote{This segregation is not problematic in itself, as it simply reflects the fact that group classes were defined based on this very gap.}. The remaining systems also exhibit a noticeably larger residual scatter.

When larger magnitude gaps involving $m_1$ are considered -- i.e.\ from $\Delta{\cal{M}}_{\rm{3-1}}$ onward -- the linear trends with $\log(f_{m_1})$ are then followed consistently across all group classes, and the variance of the residuals becomes effectively class-independent. Nevertheless, the highest values along both axes, indicative of more dynamically evolved configurations, are predominantly reached by single-BGG systems. Among these correlations, the strongest and most statistically significant are found for $\Delta{\cal{M}}_{\rm{4-1}}$, $\Delta{\cal{M}}_{\rm{5-1}}$, and $\Delta{\cal{M}}_{\rm{6-1}}$, all exhibiting Pearson coefficients in the range $r = 0.91$--$0.92$ and $p$-values below $0.001$. By contrast, as shown in the bottom-half panels of the two leftmost columns of subplots in Fig.~\ref{fig:fIGL_fBGG_mgap}, magnitude gaps defined using one of the two brightest satellites as reference instead of the main galaxy essentially segregate the groups along the $\log(f_{m_1})$ dimension, producing linear correlations with this parameter that are weak or statistically insignificant.

\begin{table*}[ht]
\centering
   \begin{threeparttable}
\caption[]{Statistics of the precursor stellar populations of the 4th-, 5th-, and 6th-ranked galaxies by group class.}
	\label{tab:merg_non}
    \centering
    \renewcommand{\arraystretch}{1.4}
    \begin{tabular}{l|c|ccc|ccc|ccc}
        \hline
         \multirow{2}{*}{\bf Group class} & {\bf Number of} & \multicolumn{3}{c|}{$\mathbf{m_4(z_{\bf fin})}$}  & \multicolumn{3}{c|}{$\mathbf{m_5(z_{\bf fin})}$} & \multicolumn{3}{c}{$\mathbf{m_6(z_{\bf fin})}$} 
        \\\cmidrule{3-11} 

                  & {\bf groups} & {\bf Ex situ}\tnote{a} & {\bf Mixed}\tnote{b} & {\bf In situ}\tnote{c} & {\bf Ex situ}\tnote{a} & {\bf Mixed}\tnote{b} & {\bf In situ}\tnote{c} & {\bf Ex situ}\tnote{a} & {\bf Mixed}\tnote{b} & {\bf In situ}\tnote{c} \\ 
        \hline
        Single-BGG & 38   & $0$   & $18$  & $20$ &  $0$  & $9$  &  $29$  & $2$  & $11$ & $25$  \\ 
        Double-BGG & 16  & $0$    & $7$   & $9$ &  $1$   & $6$  &  $9$  & $0$   & $7$ &  $9$ \\ 
        Non-BGG & 46    & $11$   & $25$   &  $10$ & $8$   & $18$ &  $20$  & $2$  & $19$ & $25$ \\
        \hline
        {\bf Total}    & 100 & 11   &  50 &  39  &  9   &   33  &  58  &  4   &   37  & 59 \\  
        \hline
        \end{tabular}
    \begin{tablenotes}
    \small 
    \item [a]$m_i(z_{\rm fin}) <0.5\ m_j(z_{\rm ini})$, with $i=4,5,6$ and $j\geq i$.
    \item [b]$0.5\ m_j(z_{\rm ini})\leq m_i(z_{\rm fin})\leq 0.9\ m_j(z_{\rm ini})$.
    \item [c]$m_i(z_{\rm fin}) > 0.9\ m_j(z_{\rm ini})$.
     \end{tablenotes}
   \end{threeparttable}
\end{table*} 

By repeating the same analysis but replacing $\log(f_{m_1})$ with $\log(f_{\rm{IGL}})$ (see the two rightmost columns of subplots in Fig.~\ref{fig:fIGL_fBGG_mgap}), we found that all correlations involving magnitude gaps defined with respect to the brightest galaxy yield Pearson coefficients in the range $r \sim 0.3$–$0.4$, indicating weaker, yet still highly significant, linear relationships. Just as occurred with the $m_1$ fraction, the correlation between $f_{\rm{IGL}}$ and $\Delta{\cal{M}}_{\rm{2-1}}$ is primarily driven by single-BGG groups, which again are the only ones that define the trend. For all other magnitude gaps of the form $\Delta{\cal{M}}_{x-1}$, with $x \in [3,6]$, we find linear relationships that are followed by groups of all classes, although with a slight tendency for single-BGG groups to show the fewest outliers. The highest correlation strengths between these gaps and the diffuse light component are obtained for $\Delta{\cal{M}}_{\rm{3-1}}$ at $r=0.38$ and $\Delta{\cal{M}}_{\rm{4-1}}$ at $r=0.37$, in both cases with highly significant $p$-values ($< 0.001$). The remaining gaps do not exhibit statistically significant correlations with $f_{\rm{IGL}}$.

Therefore, based strictly on the strength and significance of these correlations, we conclude that the magnitude gaps $\Delta{\cal{M}}_{\rm{4-1}}$ and $\Delta{\cal{M}}_{\rm{5-1}}$ are, in that order, the most suitable indicators of how far a galaxy system has progressed in its assembly, at least for groups like those in our sample with total masses on the order of a few times $10^{13}\,$\msun. This is because these two gaps (1) exhibit the strongest and most clearly defined proportional increases with both $\log(f_{m_1})$ and $\log(f_{\rm{IGL}})$, and (2) show correlations that are consistently followed by all systems, regardless of their degree of relaxation.

We also wish to remind the reader that even strong correlations between two variables should not necessarily be interpreted as evidence of a direct physical connection. Part of the observed covariation -- particularly in the relations involving $m_1$ and the magnitude gaps -- may result from their mutual dependence on the group luminosity function. Nonetheless, the existence of a consistent physical framework linking the size of the magnitude gaps to the growth of both the brightest group galaxy and the IGL fraction suggests that at least a partial physical connection is plausible. In any case, the validity of the correlations discussed here does not rely on causality, as they are used primarily as empirical diagnostics to identify which magnitude gaps most effectively trace the dynamical state of low-mass, assembling galaxy groups.

\subsection{Dependence on the assembly histories of the satellites} \label{ssec:mgap_assembly_histories}

To complement the findings above, we analysed the assembly histories of the fourth-, fifth-, and sixth-ranked galaxies in our simulated groups, with a focus on quantifying the contribution of external sources to their final stellar masses. To this end, we classified their growth histories into three distinct modes, according to the dominant origin of their assembled stellar content:
\begin{itemize}
\item Ex situ, when more than $50\%$ of the mass originates from mergers and accretion;
\item In situ, for satellites whose main progenitor at $z_{\rm ini}=3$ contributes more than $90\%$ of the final mass, indicating minimal growth during the group’s evolution;
\item Mixed, when between $10$ and $50\%$ of the mass is acquired externally.
\end{itemize}

Table~\ref{tab:merg_non} summarises the distribution of assembly modes for the fourth-, fifth-, and sixth-ranked galaxies at $z_{\rm fin} = 0$, grouped by system class. Overall, the data show that among fourth-ranked galaxies, those whose final stellar mass is significantly augmented through external contributions (i.e.\ classified as ex situ or mixed) are more common than those that remain largely unchanged since the start of the simulations (i.e.\ those in the in situ class), in a proportion of approximately $60$--$40\%$. This ratio is essentially reversed for the fifth- and sixth-ranked galaxies, where the in situ mode becomes dominant, followed by the mixed growth mode. Meanwhile, the fraction of ex situ objects decreases progressively, going from $11\%$ among fourth-ranked galaxies, to $9\%$ for fifth-ranked objects, and down to just $4\%$ for the sixth most massive group members. These findings are broadly consistent with the results of \citet{Rodriguez-Gomez+2016}, who analysed the Illustris simulation and showed that stellar mass growth in most galaxies is dominated by in situ star formation, with the two-phase ‘in situ + ex situ’ formation scenario being a good approximation only for the most massive objects \citep[see also][]{Angeloudi+2024}.

When comparing across group types, the in situ and mixed assembly modes -- particularly the former -- clearly dominate in single- and double-BGG groups, which exhibit very few 'ex situ' satellites. In single-BGG systems, the fraction of 'in situ' galaxies increases significantly among those with ranks greater than four, largely at the expense of objects with mixed assembly histories, which nonetheless hold percentages above $20\%$ in all cases examined. Double-BGG groups, on the other hand, show nearly constant proportions of satellites with in situ and mixed growing mass modes, around $60$--$40\%$, respectively\footnote{This result should be interpreted with caution due to the small number of double-BGG systems in the sample.}. In contrast, non-BGG groups display a markedly different picture: their fourth-ranked satellites predominantly follow a mixed formation pathway ($\sim 55\%$), with ex situ and in situ modes contributing nearly equally, each accounting for about $23\%$ of the cases, while among fifth- and sixth-ranked satellites the in situ and mixed modes are, in that order, the leading assembly pathways (see Table~\ref{tab:merg_non}).

\begin{figure*}
\centering
\includegraphics[width=17.5cm, height=4.5cm]{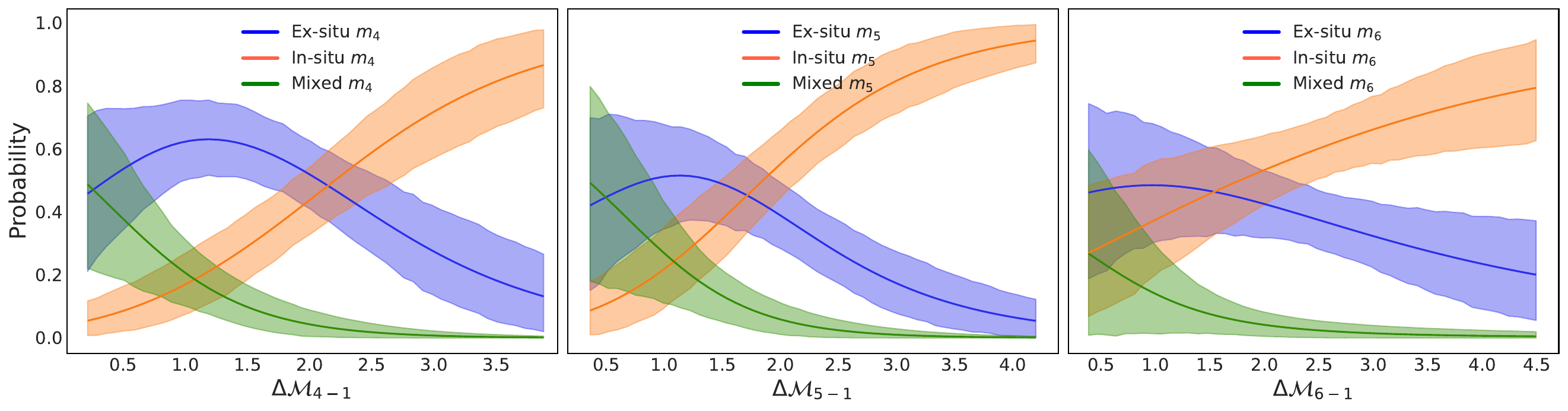}
   \caption{Fractional probabilities, accounting for the relative abundances of the assembly histories of the corresponding satellites, that the magnitude gaps $\Delta{\cal{M}}_{4-1}$ (left), $\Delta{\cal{M}}_{5-1}$ (middle), and $\Delta{\cal{M}}_{6-1}$ (right) take on a specific value at the end of the simulations. Transparent shaded bands show interquartile ranges for ex situ (blue), in situ (orange), and mixed (green) assembly modes, with solid lines indicating medians.}
   \label{fig:mgal_survival}
\end{figure*}

These results are complemented by Fig.~\ref{fig:mgal_survival}, which presents the fractional probabilities that the magnitude gaps $\Delta{\cal{M}}_{4-1}$, $\Delta{\cal{M}}_{5-1}$, and $\Delta{\cal{M}}_{6-1}$ adopt a given value at the end of the simulations, accounting for the relative abundances of assembly modes of the corresponding satellites. These probabilities, derived using a multinomial logit model \citep{agresti2012categorical}, reveal broadly consistent trends across the three magnitude gaps: smaller gap values are most frequently linked to satellites with ex situ or mixed assembly histories, while the largest gaps are almost the exclusive property of 'in situ' galaxies. The link between gap size and assembly mode is most clearly defined for gaps involving fourth-ranked satellites, and becomes somewhat less distinct for those based on lower-ranked members.

Overall, these latest findings reinforce the view that even non-first-ranked massive galaxies have a substantial probability ($\sim\!40$--$60\%$) of undergoing significant stellar mass growth through mergers and accretion during the early stages of group evolution. This, in turn, suggests that when attention is restricted to the assembly histories of the brightest satellites, the magnitude gaps $\Delta{\cal{M}}_{\rm{5-1}}$ and $\Delta{\cal{M}}_{\rm{6-1}}$ may offer more suitable alternatives than $\Delta{\cal{M}}_{\rm{4-1}}$ for gauging the dynamical state of pre-virialized groups. Conversely, if one prioritizes the sensitivity to the assembly mode of satellites, then gaps involving higher-ranked members emerge as the more effective choice.

Building on this, we provide robust new evidence that certain magnitude gaps -- particularly those involving the fourth-, fifth-, and sixth-brightest satellites -- serve as more reliable proxies of group formation time than the traditional $\Delta{\cal{M}}_{\rm{2-1}}$. In this respect, our results complement and extend earlier theoretical studies, like those by \citet{Dariush+2010}, who were the first to highlight the advantages of the $\Delta{\cal{M}}_{\rm{4-1}}$ gap for identifying early-formed systems, and by \citet{Farahi+2020}, who later demonstrated its effectiveness in reducing scatter in halo property-mass relations. Consistent with our findings, \citeauthor{Farahi+2020} further reported an anti-correlation between formation time and the number of surviving satellites, which they interpreted as a natural outcome of hierarchical assembly. Our results are likewise supported by the recent observational analysis of \citet{Golden-Marx+2025}, who, using a large cluster sample, showed that $\Delta{\cal{M}}_{\rm{4-1}}$ correlates with both the ICL and the combined BCG + ICL mass fractions, while confirming its role in reducing the scatter in the stellar mass-halo mass relation. Collectively, all these findings reinforce the view that wider magnitude gaps provide more effective diagnostics of hierarchical growth, central-galaxy assembly, and diffuse light build-up in complex galaxy systems.

That being said, our analyses also indicate that the three magnitude gaps identified as the most suitable differ only modestly in their overall effectiveness as indicators of group dynamical state. This suggests that none of these metrics is universally superior to the others, so the optimal choice will ultimately depend on additional factors beyond the scope of the present study. For instance, in observational work, a key concern is that the lower the rank of the secondary galaxy used to compute the gap, the more susceptible the measurement becomes to contamination by interlopers (i.e.\ non-member galaxies). This issue is further compounded in shallow surveys, where fainter satellites are more likely to approach the detection limit or fall below the survey’s completeness threshold. Similarly, in poor groups with few genuine members, magnitude gaps involving lower-ranked satellites may become unreliable or altogether unusable.

Similarly,  there are also compelling reasons for favouring magnitude gaps defined using fainter secondary galaxies, as these tend to be less sensitive to the reordering of ranks caused by the occasional non-detection of some of the brightest members, making them potentially more robust under incomplete or uncertain sampling conditions.

\section{Summary and conclusions}\label{sec:conclusions}

This work builds upon the same suite of 100 cosmologically motivated, controlled numerical simulations employed in \citetalias{paperI}, which follow the early gravitational collapse of low-mass galaxy groups ($\sim\!1$--$5\,\times\;\!10^{13}\,$\msun). Here, we focused on tracing the evolutionary histories of several widely used indicators of group dynamical relaxation, with particular attention to the secular evolution of the diffuse IGL and the top-ranked galaxies, as well as identifying the mass distributions of the progenitor galaxies that contribute to the final masses and fractional abundances of these components. Our analysis concentrated on the evolutionary stage spanning from just before the groups’ turnaround epoch to the completion of their initial non-linear infall. As in \citetalias{paperI}, we adopted a novel approach consisting on comparing outcomes across three group classes defined by the number of dominant galaxies present at the end of the simulated period, namely, single-BGG, double-BGG, and non-BGG systems. Special emphasis was placed on examining the correlations between various magnitude gaps defined from the brightest group members and the stellar mass fractions of the first-ranked galaxy and the IGL, the two components most directly impacted by the hierarchical assembly of the groups. We also traced the assembly histories of the fourth-, fifth-, and sixth-ranked group galaxies. These analyses were used to identify which magnitude gaps serve as the most reliable proxies for dynamical age in these systems. Our main findings are summarised below:

   \begin{itemize} 
    
    \item Examination of the secular evolution of the surviving galaxy fraction during the first infall epoch of groups indicates that, over the last $\sim 9.3$ Gyr, the merger activity can be reasonably approximated as a stochastic Poisson process, characterized by a steady merger rate. The only exception occurs in single-BGG systems, which exhibit a more gradual decline in group membership during the final $\sim 3$ Gyr of the runs.

    \item Single-BGG groups are also characterized by both smaller pairwise intergalactic separations and smaller positional offsets between the first-ranked galaxy and the group’s centre of mass. Additionally, they exhibit a faster progression towards a more ordered distribution of galaxy velocity dispersions compared to the double- and non-BGG group classes. 
    
    \item In a similar vein, we find that the total amount of ex situ IGL and its growth rate are broadly insensitive to group class and total mass, with single-BGG groups being the only systems to exhibit a mildly distinctive behaviour, showing a tendency to produce slightly higher IGL fractions over cosmic time than the other group types.
    
    \item The evolutionary patterns of the various dynamical state indicators investigated support the notion that our BGG-based classification effectively correlates with the dynamical age of pre-virialised groups, with single-BGG systems evolving more rapidly towards dynamical maturity, whereas double- and, especially, non-BGG groups experience a more gradual and less efficient dynamical evolution.

    \item We present further evidence that, despite their intertwined origins, the first-ranked group galaxies and the IGL experience distinct growth histories, shaped by different sets of progenitors. In particular, we find that the relevant progenitors of the IGL differ markedly from those contributing to the main group galaxy, both in typical mass and in overall mass range, with the IGL being assembled from a broader and systematically lower-mass galaxy population. 

    \item We also find that the weak yet statistically significant negative linear correlation between the logarithms of the IGL mass fraction and the mass-weighted velocity dispersion of group galaxies observed at the end of our simulations emerges during the final $\sim 3.5$ Gyr of the runs, coinciding with the late stages of gravitational collapse and the main phase of IGL assembly.

    \item In single- and double-BGG groups, the IGL is predominantly assembled from more massive progenitors, with over $50\%$ of its principal contributors ranking among the ten most massive galaxies at the start of the simulations. In contrast, the IGL in non-BGG systems originates from a broader range of progenitor masses, including a significant contribution from low-mass galaxies. In these systems, approximately one-third of the total IGL mass is sourced from small objects that individually account for less than $5\%$ of this component.

    \item Our simulations provide compelling evidence that not only the first-ranked galaxies but also the three most massive satellites in galaxy groups typically participate -- albeit to a lesser extent -- in the hierarchical assembly of these systems. For these major galaxies, stellar mass growth is predominantly driven by ex situ accretion, with the fraction of objects that increase their mass through mergers ranging from $100\%$ for the centrals to about $63\%$ for the fourth-ranked members.
    
    \item Based on the strength and significance of their correlations with the stellar mass fractions in both the main galaxy and the IGL, we identify the magnitude gap $\Delta{\cal{M}}_{\rm{4-1}}$, followed by $\Delta{\cal{M}}_{\rm{5-1}}$, as more suitable proxies for the dynamical state of unrelaxed groups than the traditional $\Delta{\cal{M}}_{\rm{2-1}}$ gap -- at least for low-mass systems such as those analysed in this work. However, when the satellite assembly histories are considered instead, the preference shifts towards gaps involving the fifth and sixth-ranked galaxies. In any case, the differences in performance among the $\Delta{\cal{M}}_{\rm{4-1}}$, $\Delta{\cal{M}}_{\rm{5-1}}$, and $\Delta{\cal{M}}_{\rm{6-1}}$ gaps are not substantial enough to single out a clear best choice. So, the selection of one of these dynamical age indicators over the others will ultimately depend on the specific goals and characteristics of the study in which it is employed.  

   \end{itemize}

\begin{acknowledgements}
BBW acknowledges support by the PRE2020-093715 pre-doctoral grant from the SEV–2017–0709 project. This work also benefited from the support provided by the Spanish state agency MCIN/AEI/10.13039/501100011033 and by 'ERDF A way of making Europe' funds through research grants PID2022-140871NB-C21 and PID2022-140871NB-C22. The MCIN/AEI/10.13039/501100011033 has also provided additional support through the Centre of Excellence Severo Ochoa's award for the Instituto de Astrof\'\i sica de Andalucía under contract CEX2021-001131-S and the Centre of Excellence Mar\'\i a de Maeztu's award for the Institut de Ci\`encies del Cosmos at the Universitat de Barcelona under contracts CEX2019–000918–M and CEX2024-001451-M. 
\end{acknowledgements}

\bibliographystyle{aa} 
\bibliography{biblio}

\begin{appendix}
\onecolumn
\section{Correlations between stellar mass fractions and magnitude gaps}\label{app:A}
This appendix presents the correlations between the logarithmic stellar mass fractions in $m_1$ and in the IGL and several definitions of the magnitude gap, complementing the analysis discussed in Sect.~\ref{ssec:mgap_correlations}.

\begin{figure*}[ht!]
\centering
 \includegraphics[width=9cm, height=20cm]{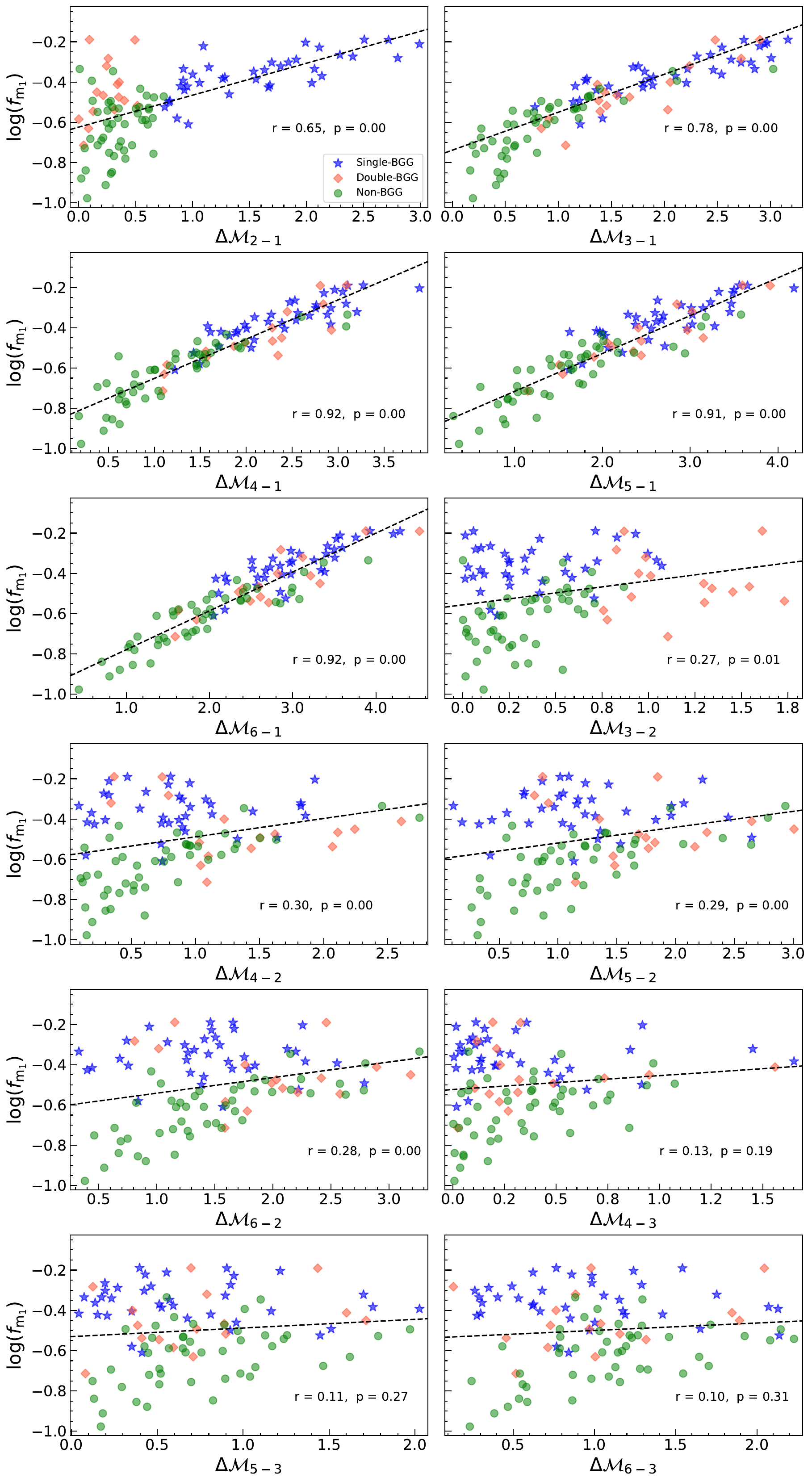}
 \includegraphics[width=9cm, height=20cm]{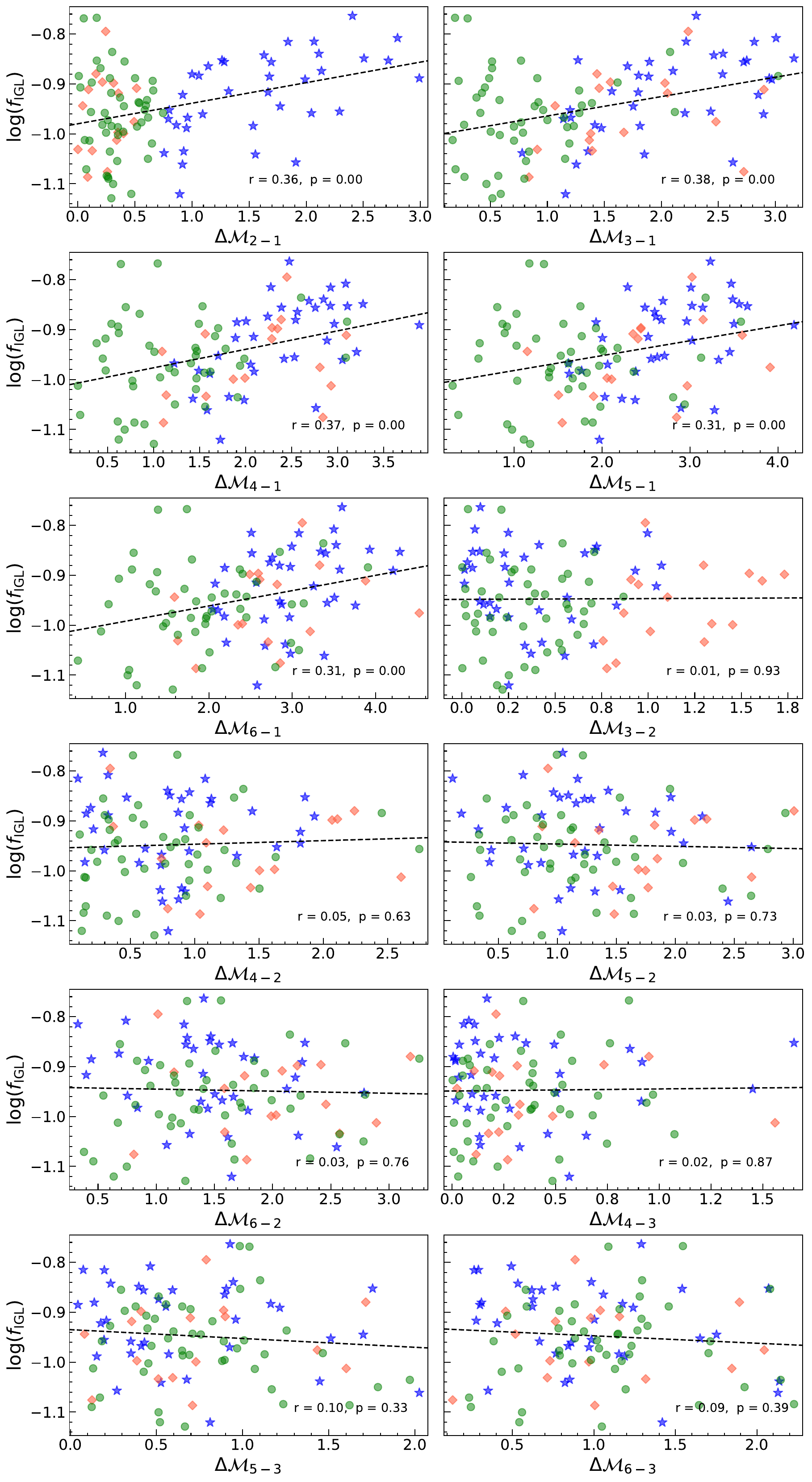}
    \caption{Relations between the logarithm of the mass fractions in  $m_1$ (two left columns) and the IGL (two right columns) at the end of the simulations and various magnitude gaps defined from the first six ranked galaxies in the groups. Each panel displays the Pearson $r$ coefficient and associated $p$-value, with dotted black lines indicating the best linear fit. Weak or statistically insignificant linear correlations should not be taken as evidence of a physical link. The colours and shapes of data points identifying group classes follow the same scheme used in Fig.~\ref{fig:m1_mtot}.}
    \label{fig:fIGL_fBGG_mgap}
\end{figure*}

\end{appendix}
\end{document}